\documentclass[12pt]{iopart}

\pdfoutput=1

\usepackage{bm}
\usepackage{braket}
\usepackage{amssymb}
\usepackage{amsfonts}
\usepackage{amsmath}
\usepackage{graphicx}
\usepackage{bm}
\usepackage{xcolor}
\usepackage{hyperref}
\usepackage[toc,title]{appendix}

\def\be{\begin{equation}}
\def\ee{\end{equation}}

\def\bea{\begin{eqnarray}}
\def\eea{\end{eqnarray}}

\def\Tr{{\rm Tr}}

\begin{document}

\title[Entanglement negativity after a global quench]{Entanglement negativity after a global quantum quench}

\vspace{.5cm}

\author{Andrea Coser$^1$, Erik Tonni$^1$ and Pasquale Calabrese$^{1,2}$}
\address{$^1$\,SISSA and INFN, via Bonomea 265, 34136 Trieste, Italy. \\}
\address{$^2$\,Dipartimento di Fisica dell'Universit\`a di Pisa and INFN,
             Pisa, Italy.}

\vspace{.5cm}

\begin{abstract}

We study the time evolution of the logarithmic negativity after a global quantum quench.
In a 1+1 dimensional conformal invariant field theory, we consider the negativity between two  
intervals which can be either adjacent or disjoint. 
We show that the negativity follows the quasi-particle interpretation for the spreading of entanglement. 
We check and generalise our findings with a systematic analysis of the negativity after a quantum quench 
in the harmonic chain, highlighting two peculiar lattice effects: the late birth and the sudden death of entanglement.

\end{abstract}

\maketitle


\section{Introduction}
\label{sec intro}

The non-equilibrium dynamics of isolated quantum systems is one of the most active research 
area of the last years. 
In a global quantum quench, a system is initially prepared in the ground state of a translationally invariant Hamiltonian $H_0$
and it is then left evolving with another translationally invariant Hamiltonian $H$ differing from $H_0$ for an experimentally 
tuneable parameter. 
Key questions in quench dynamics are whether the system reaches for long time a stationary state, 
how to characterise  it from first principles, and how this steady state is approached in time (see e.g. Refs.~\cite{silva,efg-14} for reviews). 

Nowadays a number of advanced analytical and numerical techniques have been developed to study the quench dynamics  
for a variety of different situations and realistic models \cite{cc-06,ce-13,a-12,tdmrg,iTEBD,s-11}. 
However, many insights on these non-equilibrium dynamics came from the study of oversimplified theories such as 1+1 dimensional 
conformal field theory (CFT).
Indeed, phenomena like the light-cone spreading of correlations \cite{cc-06,cc-07-quench}, the linear increase of entanglement 
entropy \cite{cc-05-quench} and the structure of revivals in finite systems \cite{c-14} have been first discovered in CFT, later generalised to more realistic models and even verified in experiments (see \cite{cetal-12} for the experimental measure of the light-cone spreading of correlations).  

The main goal of this paper is to shed some light on the time evolution of the entanglement {\it between} two different regions in an 
extended system following a quantum quench. 
We will quantify this entanglement by means of logarithmic negativity for which a general quantum field theory 
approach has been recently developed \cite{us-letter,us-long,us-neg-T}. 
We consider this problem in the framework of CFT, closely following the approach introduced in 
Refs.~\cite{cc-05-quench, cc-07-quench} for the time evolution of entanglement entropy and correlations.
In order to understand the generality and the limits of this approach, we parallel the analytic CFT calculations with some 
exact numerical computations for the harmonic chain.
The non-equilibrium evolution of the negativity in CFT, but for different quench protocols,  has also
been considered in Refs.~~\cite{ez-14,d-14}. 

\subsection{Quench protocol}
The system is prepared in the ground state $| \psi_0 \rangle$ of the Hamiltonian $H_0$. 
The quantum quench consists in a sudden (instantaneous) change of a parameter in the Hamiltonian $H_0 \rightarrow H$ 
at a given time that we set as $t=0$. Thus, the unitary evolution of $| \psi_0 \rangle$ is 
\be
\label{state t>0}
| \psi(t) \rangle =  e^{-\textrm{i} H t} \, | \psi_0 \rangle\,.
\ee
The density matrix associated to this pure state is $\rho(t) = | \psi(t) \rangle  \langle \psi(t) |$.
In 1+1-dimensional CFT, the calculations become manageable when $|\psi_0\rangle$ is a boundary conformal state as we will 
explain in what follows. 
In this approach, analytical results have been obtained for the entanglement entropy of a single and more intervals \cite{cc-05-quench}, 
for correlation functions of primary operators \cite{cc-06,cc-07-quench}, and a few  other quantities \cite{sc-08,gs-12,mg-14,nnt-14}.
We will not be interested here in the time evolution of the entanglement after a local quench, a subject which has been 
considered instead in Refs.~\cite{cc-07l,ep-08,ds-11,ep-12,c-12,ab-14}, but only for the entanglement entropies.

\subsection{Quantities of interest}

\begin{figure}[t]
\vspace{.4cm}
\begin{center}
\includegraphics[width=.9\textwidth]{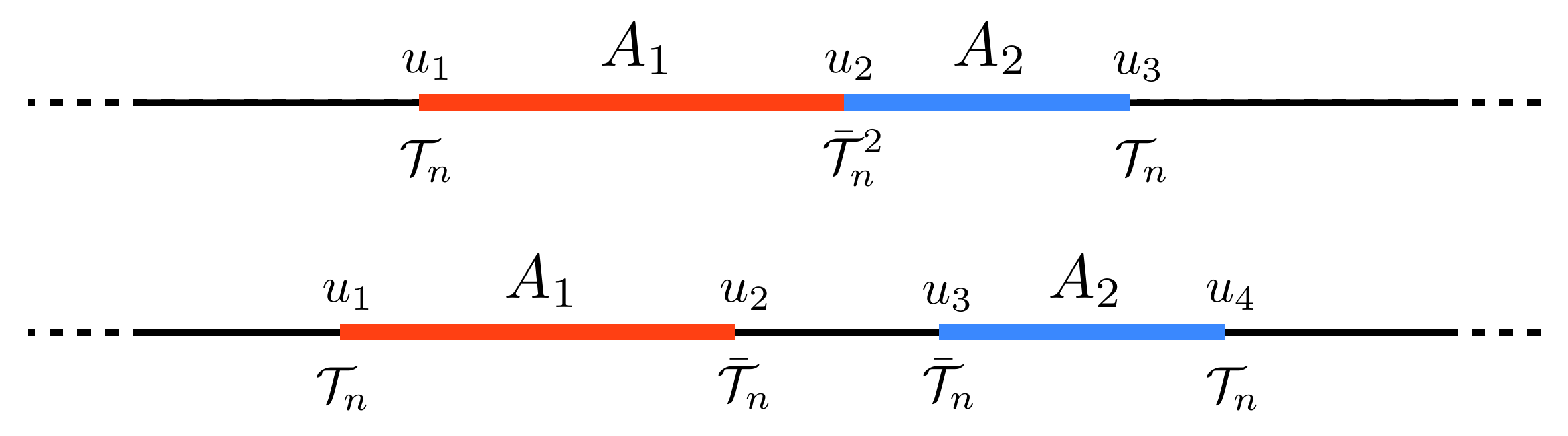}
\end{center}
\vspace{-.3cm}
\caption{Configurations of two intervals in the infinite line that we consider: adjacent intervals (top) and disjoint intervals (bottom). 
}
\label{fig line intervals}
\end{figure}

As we anticipated, we are interested here in the entanglement between two different regions, which in the case of a one dimensional 
system are two intervals, either adjacent or disjoint, as shown in Fig.~\ref{fig line intervals}.
In order to define the entanglement, we should first introduce the reduced density matrix of the part $A$ of the 
system as $\rho_A={\rm Tr}_B \rho(t)$, where we traced over the degrees of freedom in $B$, which is the complement of $A$.

The entanglement of a bipartite system can be quantified by the entanglement entropy \cite{rev}
\be 
S_A= - {\rm Tr}\,\rho_A \ln \rho_A\,,
\label{Sdef}
\ee
or alternatively by the R\'enyi entanglement entropies 
\be
S_A^{(n)} = \frac1{1-n} \ln {\rm Tr}  \rho_A^n\,.
\label{SAn}
\ee
The limit $n\to1$ of $S_A^{(n)} $ gives $S_A$, but $S_A^{(n)}$ contain much more information, since one can extract the full spectrum of $\rho_A$ from them \cite{cl-08}.
Notice that, when $A=A_1\cup A_2$ is composed of two disjoint regions $A_1$ and $A_2$, the entanglement entropy quantifies only the entanglement between $A$ and the remainder of the system $B$, but not the entanglement between 
$A_1$ and $A_2$. 
In this case one can introduce  the mutual information
\be
\label{MI def}
I_{A_1,A_2}\equiv\, S_{A_1} + S_{A_2} - S_{A_1 \cup A_2}\,,
\ee
and, analogously, the R\'enyi  mutual information 
\be
\label{renyi MI def}
I^{(n)}_{A_1,A_2}\equiv\,
 \frac{1}{n-1}
\ln \left( \frac{\Tr \rho_A^n}{\Tr \rho_{A_1}^n  \Tr \rho_{A_2}^n} \right) .
\ee
However, these are not measures of the entanglement between $A_1$ and $A_2$, but quantify the amount of 
{\it global correlations between the two subsystems}, see e.g. \cite{MI}.

A proper measure of entanglement in a bipartite mixed state  is the negativity \cite{vw-01,horodecki-98-neg,eisert-01}. 
In order to define it, one first introduces the partial transpose  with respect to the $A_2$'s degrees of freedom  as
\be 
\langle e_i^{(1)} e_j^{(2)}|\rho^{T_2}_{A_1\cup A_2}|e_k^{(1)} e_l^{(2)}\rangle=
\langle e_i^{(1)} e_l^{(2)}|\rho_{A_1\cup A_2}| e^{(1)}_k e^{(2)}_j\rangle\, ,
\label{rhoAT2def}
\ee
and then the  {\it logarithmic negativity} as
\be
{\cal E}_{A_1,A_2}\equiv\ln ||\rho^{T_2}_{A_1\cup A_2}||=\ln \Tr |\rho^{T_2}_{A_1\cup A_2}|\,,
\label{negdef}
\ee
where $||M|| \equiv \sum_i |\lambda_i|$ is the trace-norm of the hermitian matrix $M$, defined through its eigenvalues $\lambda_i$.
Notice that the negativity is symmetric for exchange of $A_1$ and $A_2$, as any good measure of the relative 
entanglement should be. 
While the negativity was introduced long time ago, only recently it has become a practical and useful tool for the study of
many-body quantum systems \cite{Neg1,Neg2,Neg3,sod,kor,hv-14,us-letter,us-long,us-neg-T}.

\subsection{Organisation of the manuscript}

The main goal of this paper is to study the time dependence following a quantum quench of the negativity  ${\cal E}_{A_1,A_2}$ and 
compare it with the mutual information. 
The paper is organised as follows. 
In Sec.~\ref{sec EE and MI} we review the path integral approach to the 
quantum quench problem \cite{cc-05-quench,cc-06,cc-07-quench} and the known results from the time evolution 
of the entanglement entropy and mutual information in a CFT. 
In Sec.~\ref{sec ent neg} we apply this formalism to the calculation of the negativity of two disjoint intervals after a quench to a CFT. 
In Sec.~\ref{QP} we report the quasi-particle picture for the spreading of correlations and entanglement and we argue 
that it is valid also for the negativity. 
In Sec.~\ref{app hc} we report numerical calculation for the entanglement entropy, mutual information, and entanglement negativity 
for a quench of the frequency (mass) in the harmonic chain. 
We show that the results are in qualitative agreement with the CFT predictions and the differences are understood in 
terms of the effect of slow quasi-particles.
Finally in Sec.~\ref{concl} we draw our conclusions and we discuss some open problems.

\section{Entanglement entropies and mutual information}
\label{sec EE and MI}

In this section we briefly review the imaginary time formalism for the description of quenches in CFTs developed in 
Refs.~\cite{cc-05-quench,cc-06,cc-07-quench}.
In particular in Ref.~\cite{cc-05-quench}, the entanglement entropies of an arbitrary number of disjoint intervals
have been already derived.
It is however useful to recall how this has been done, in order to set up the calculation and notations for the negativity. 

\subsection{The path integral approach to quenches}

The expectation value of a product of equal-time local operators in the time dependent state (\ref{state t>0}) can be written as \cite{cc-06}
\be
\langle {\cal O}(t,\{ { r}_i\})\rangle= Z^{-1} \langle \psi_0 |
e^{\textrm{i} H t-\tau_0 H} {\cal O}(\{ { r}_i\}) e^{-\textrm{i} H t-\tau_0 H}| \psi_0 \rangle\, , 
\label{Oexp}
\ee
where two damping factors $e^{-\tau_0 H}$ have been added in such a way
as the path integral representation of this expectation value is convergent. 
The normalisation factor is $Z=\langle\psi_0|e^{-2\tau_0 H}|\psi_0\rangle$. 
Following Ref.~\cite{cc-06}, Eq.~(\ref{Oexp}) may be represented by a path integral in imaginary time $\tau$
\be
\label{pi} \frac1Z\int[d\phi(r,\tau)]{\cal O}(\{ r_i\},0) \,e^{-\int_{\tau_1}^{\tau_2}Ld\tau} 
\langle\psi_0|\phi( r,\tau_2)\rangle \langle\phi( r,\tau_1)|\psi_0\rangle \, ,
\ee
where $L$ is the (euclidean) Lagrangian corresponding to the dynamics of $H$. 
In order to match the expectation value (\ref{pi}) with the starting formula (\ref{Oexp}) we need to identify $\tau_1=-\tau_0-\textrm{i}t$ and $\tau_2=\tau_0-\textrm{i}t$. 
To further simplify the calculation and following Ref.~\cite{cc-06}, we consider
the equivalent strip geometry between $\tau=0$ and $\tau=2\tau_0$, with
$\cal O$ inserted at $\tau=\tau_0+\textrm{i}t$.
The main idea of Refs.~\cite{cc-05-quench,cc-06} is to make the calculation considering $\tau$
real and only at the end of the computation to analytically continue it to the actual complex value  $\tau=\tau_0+\textrm{i}t$. 

Eq.~(\ref{pi}) has the form of the equilibrium expectation value in a strip of width $2\tau_0$ with particular boundary conditions. 
The above expression is valid for an arbitrary field theory, but it is practically computable in the case we are interested in, i.e.  
a conformal invariant Hamiltonian.
As detailed in Ref.~\cite{cc-06} for a CFT, in the limit when $t$ and the separations $| r_i- r_j|$ are much larger than the
microscopic length and time scales, we can replace the boundary condition $|\psi_0\rangle$ with a boundary conformal state   
$|\psi_0^*\rangle$ to which $|\psi_0\rangle$ flow under the renormalization group flow. 
Within this approach $\tau_0$ is  identified with the correlation length (inverse mass) of the initial state and the 
predictions made with this approach are expected to be valid only in the regime $t, |r_i-r_j|\gg \tau_0$.
The generalisation of this approach to some other initial conditions (both in one and higher dimensions) 
can be found in Refs.~\cite{sc-08,nnt-14,chd-08,cardy-talk-ggi,sc-10,gc-11,stm-14}. 

Before reporting the explicit results and technicalities for the entanglement entropies, 
it is worth spending few words on the regime of applicability of the above approach that often in the literature has been taken 
much beyond its scope, especially when comparing with results in lattice models.  
First of all, in a CFT all the quasi-particle excitations move with the same speed which here has been fixed to unity. 
This is not the case for a critical  model even if its low-energy physics is described by a CFT.
Indeed, while for small momenta $k$ the dispersion relation $\epsilon_k$ has a CFT form $\epsilon_k\sim v |k|$, 
for larger values of the momentum $k$ it becomes a non-trivial function.
When performing a global quench, we always inject a large amount of energy into the system (unless we perform an infinitesimal quench) 
which populates also high-energy modes having a non-conformal scaling. 
Also the identification of $\tau_0$ should be handled with a lot of care. 
Indeed, for small initial correlation length $\xi_0$ we have $\tau_0\sim \xi_0$, but this relation should be seen only  
as an effective scaling for small $\xi_0$ in the continuum theory. 
However, in a given lattice model we need to have $\xi_0\gg a$ in order to be in the field theory scaling. 
Thus there is a competition between two different effects, which makes $\tau_0$ a non-univocally defined quantity. 
However, this is not our main interest in the following and, when comparing with the numerical results coming from the 
harmonic chain, we will simply limit ourself to use $\tau_0$ as a phenomenological fitting parameter which can depend
also on the considered observable (as already noticed a few times in the literature \cite{sc-08,cardy-talk-ggi,fm-10}). 

\subsection{The entanglement entropy}

Let us consider a subsystem $A=\cup_{i=1}^N A_i$ composed by $N$ disjoint intervals $A_i =[u_{2i-1}, u_{2i}]$ on the infinite line. 
We are interested in the time-dependent R\'enyi entropies $S^{(n)}_A(t)$ as defined in Eq.~(\ref{SAn}) and in the entanglement entropy obtained as a replica limit. 
Given that $ \Tr \rho_A^n$ is equivalent to a $2N$-point  function of twist fields \cite{cc-04,cc-rev,ccd-09}, we have that the 
desired imaginary-time expectation value is 
\be
\label{renyi corr strip}
 \Tr \rho_A^n
 =
\langle\, \prod_{i=1}^N 
\mathcal{T}_n (w_{2i-1}) 
\bar{\mathcal{T}}_n(w_{2i}) \,\rangle_{\rm strip} \,,
\qquad
w_i= u_i + \textrm{i} \tau \,,
\ee
where we denoted by  $w= u + \textrm{i} \tau $  the complex coordinate on the strip ($u \in \mathbb{R}$ and $0< \tau< 2\tau_0$).
The twist fields $\mathcal{T}_n$ and $\bar{\mathcal{T}}_n$ behave under conformal transformation as primary operators whose 
scaling dimensions are given by \cite{cc-04, Knizhnik-87}
\be
\label{Delta_n def}
\Delta_n =   \frac{c}{12}\left( n- \frac{1}{n} \right) .
\ee

The expectation values on the strip of width $2\tau_0$ can be obtained
by employing the conformal map $z = e^{\pi w / (2\tau_0)}$, which maps the strip to the upper half plane (UHP) 
parameterised by complex coordinate $z$.  Eq.~(\ref{renyi corr strip}) can be then written as
\be
\label{renyi cft strip2uhp}
 \Tr \rho_A^n
=
\bigg[
\bigg(\frac{\pi}{2\tau_0} \bigg)^{2N}
\prod_{i=1}^N \big| z_{2i-1}  z_{2i} \big|
\bigg]^{\Delta_n}
\langle\, \prod_{i=1}^N 
\mathcal{T}_n (z_{2i-1}) 
\bar{\mathcal{T}}_n(z_{2i}) \,\rangle_{\rm UHP}\,,
\ee
where $\langle \dots \rangle_{\rm UHP} $ are correlators on the UHP and $z_j \equiv e^{\pi w_j / (2\tau_0)} $.

The $2N$-point function of twist fields on the upper half plane occurring in (\ref{renyi cft strip2uhp}) can be written as 
\be
\label{corr uhp renyi}
\langle\, \prod_{i=1}^N 
\mathcal{T}_n (z_{2i-1}) 
\bar{\mathcal{T}}_n(z_{2i}) \,\rangle_{\rm UHP} 
 \,=\,  
\frac{c_n^N}{ \prod_{a=1}^{2N} | z_a - \bar{z}_a|^{\Delta_n}}
\left(\frac{\prod_{j<k}^N \eta_{2k,2j}\, \eta_{2k-1,2j-1}}{\prod_{j,k}\eta_{2j-1,2k}}\right)^{\Delta_n}{\cal F}_{N,n}(\{\eta_{j,k}\})
\,,
\ee
where $\eta_{i,j} $ are the $\binom{2N}{2}$ cross ratios that can be constructed from the $2N$ endpoints $z_j$ (and their images $\bar z_j$)
of the $N$ intervals in the UHP as follows
\be
\label{eta cross ratio uhp}
\eta_{i,j} 
\equiv
\frac{(z_i - z_j)(\bar{z}_i - \bar{z}_j)}{(z_i - \bar{z}_j)(\bar{z}_i - z_j)}\,.
\ee
The function $ \mathcal{F}_{N,n}(\{\eta_{j,k}\})$ in (\ref{corr uhp renyi}) depends on the full operator content of the model and its
computation is a very difficult task, even for simple models (see  
Refs.~\cite{cd-08,fps-08,cct-09,cct-11,ip-09,c-10,ch-09,atc-10,atc-11,h-10,rg-12,f-12,cz-13,ctt-14} for some specific cases in the bulk case,
the references in \cite{hol} for the holographic approach to the same problem, and \cite{c-13b,ch-09b,s-12, sch-14} for some higher dimensional 
field theoretical computations).

There is, however,  some degree of arbitrariness in the way we wrote Eq.~(\ref{corr uhp renyi}) since the product over the cross-ratios
could be absorbed fully or partially in the function ${\cal F}_{N,n}$.
However,  writing it in the above form has the advantage to display the limiting behaviour for $\eta_{i,j}\to 0$ and 
$\eta_{i,j}\to 1$. Indeed, by employing the operator product expansion (OPE) (in the sense of Ref.~\cite{cct-11})
\be
\label{ope TTbar}
\mathcal{T}_n(z) \bar{\mathcal{T}}_n(w) 
=
\frac{c_n}{|z-w|^{2\Delta_n}}\, \mathbb{I} + \dots\,,
\qquad
w \rightarrow z\,,
\ee
it is easy to show that, in both limits $\eta_{i,j}\to 0$ and $\eta_{i,j}\to 1$, the leading power-law behaviour is fully encoded
in the prefactor and the function ${\cal F}_{N,n}$ is just a constant.
As we will see below, for the real time behaviour of the entanglement entropy only these two limits of the various four-point
ratios matter \cite{cc-05-quench,cc-06} and consequently we do not have to worry about the precise value of  
the function ${\cal F}_{N,n}$.


Plugging (\ref{corr uhp renyi}) into (\ref{renyi cft strip2uhp}), we find
\be
\label{renyi as  corr strip}
 \Tr \rho_A^n
\,=\,
c_n^N \Bigg[
\left( \frac{\pi}{2\tau_0} \right)^{2N}
\prod_{a=1}^{2N} \bigg| \frac{ z_a}{z_a - \bar{z}_a} \bigg| 
\frac{\prod_{j<k}^N \eta_{2k,2j} \, \eta_{2k-1,2j-1}}{\prod_{j,k}  \eta_{2j-1,2k}}\Bigg]^{\Delta_n}{\cal F}_{N,n}(\{\eta_{j,k}\}) \,.
\ee

What still remains to be done is to 
write the r.h.s. of (\ref{renyi as  corr strip}) in terms of the coordinates on the strip.
The $a$-th term of the product in (\ref{renyi as  corr strip}) is simply 
\be
\frac{|z_a|}{|z_a - \bar{z}_a|} = \frac{1}{ | 2  \sin [ \pi \tau/(2\tau_0)] |} \, ,
\label{a-term}
\ee 
independently of $u_a$.
For the cross ratios (\ref{eta cross ratio uhp}) we have that
\be
\label{eta strip coords}
\eta_{i,j} \,=\, \frac{2 \sinh^2\big(\tfrac{\pi(u_i - u_j)}{4\tau_0}\big)}{
\cosh\big(\tfrac{\pi(u_i - u_j)}{2\tau_0}\big) - \cos\big(\tfrac{\pi \tau}{\tau_0}\big)}\,.
\ee

This concludes the calculation on the strip of width $2\tau_0$. At this point, to obtain the real time evolution after a quench,
we should analytically continue the parameter $\tau$ to the complex value
\be
\label{analytic cont}
\tau = \tau_0  + \textrm{i} \,t\,,
\ee
with $t \gg \tau_0$, as explained above.
In this regime, the $a$-th term (\ref{a-term}) gives $|z_a|/| z_a - \bar{z}_a|=  e^{-\frac{\pi}{2\tau_0} t} + \dots$\,.
As for the cross ratio (\ref{eta strip coords}), when $t \gg \tau_0$ and $|u_i - u_j| \gg \tau_0$,  it becomes
\be
\label{etaij small tau0}
\eta_{i,j} 
\,=\, 
\frac{e^{\pi |u_i - u_j|/(2\tau_0)}}{
e^{\pi |u_i - u_j|/(2\tau_0)} + e^{\pi t/\tau_0}}\,,
\ee
where $\eta_{i,j} \in [0,1]$.
In the limit $\tau_0 \rightarrow 0$, for the ratio (\ref{etaij small tau0}) we have $\eta_{i,j} \to 0$ for  $ t> |u_i -u_j|/2$ 
and $\eta_{i,j} \to 1$ for $ |u_i -u_j|/2 > t$. 
However, we should keep the leading behaviour for $\eta_{i,j} \to 0$ and so we find 
useful to write it as 
\be
\label{log eta time regimes}
\ln (\eta_{i,j}) \to \frac{\pi}{\tau_0}  q(t,|u_i - u_j|) \,,
\qquad {\rm for} \quad t,|u_i - u_j|  \gg \tau_0  \, ,
\ee
where 
\be
\label{qdef}
q(t,\ell) \,\equiv\, \frac{\ell}{2} -  \textrm{max}(t, \ell/2)
\,=\,
\left\{\begin{array}{ll}
0 \hspace{2cm}& t <\ell/2\,, \\
\ell/2-t  & t >\ell/2\,.
\end{array}\right.
\ee

The time evolution of the R\'enyi entanglement entropies is then obtained by plugging the above analytic continuations 
to real time in Eq.~(\ref{renyi as  corr strip}), leading to \cite{cc-05-quench}
\begin{multline}
S_A^{(n)}=\frac{c \pi(n+1)}{12\tau_0 n }\left[N t+\sum_{j,k=1}^N q(t,|u_{2j-1}-u_{2k}|) \right. \\ \left. -
\sum_{1<j<k<N}q(t,|u_{2j}-u_{2k}|)+q(t,|u_{2j-1}-u_{2k-1}|)
\right].
\label{Sngen}
\end{multline}
In the above equation a piece-wise constant (in time) term coming from the function ${\cal F}_{N,n}$ and for the various non-universal 
prefactors has been {\it intentionally} dropped because it has no physical meaning. 
Indeed the above formula describes only the leading term in the so-called {\it space-time scaling limit} \cite{fc-08,CEF,CEFI} 
which corresponds to the limit $t\to\infty$, $|u_i-u_k|\to\infty$ with all the ratios fixed. 
The term we dropped is just one of the corrections to this leading behaviour and it has no meaning to consider it without 
taking into account all other corrections at the same order, such the dependence on the details of the initial state (which in this approach 
have been over-simplistically absorbed in the parameter $\tau_0$).

Let us now specialise Eq.~(\ref{Sngen}) to the case of one interval of length $u_2-u_1=\ell$ 
and $n=1$ obtaining the well-known formula \cite{cc-05-quench}
\be
\label{SA one interval}
S_A = \frac{\pi c}{6\tau_0}\big[t + q(t, \ell) \big]= 
\left\{\begin{array}{ll}\displaystyle 
\frac{\pi c}{6\tau_0}t \hspace{2cm}& t <\ell/2\,, \\ \\ \displaystyle 
\frac{\pi c}{12\tau_0} \ell & t >\ell/2\,,
\end{array}\right.
\ee
i.e. the entanglement entropy grows linearly for $t<\ell/2$ and then saturates to an extensive value in the 
subsystem length  $\ell$.
Notice that the large time value of the entanglement entropy is the same as the thermodynamic entropy of 
a CFT at large finite temperature $T=4\tau_0$. This fact indeed holds for all local observable leading to the 
remarkable phenomenon of CFT thermalisation \cite{cc-06} (see also \cite{hol-therm} for the holographic 
version of this phenomenon in arbitrary dimension).
However, this is a specificity of the uncorrelated initial state we are considering and it has been shown that
even an irrelevant boundary perturbation destroys it leading, for large time, to a generalised Gibbs ensemble 
where all the CFT constants of motion enter \cite{cardy-talk-ggi}. 
The discussion of this issue is however far beyond the goals of this paper.

In the case of two intervals $A_1$ and $A_2$ (the geometry depicted in Fig.~\ref{fig line intervals} with $u_2-u_1$ and $u_4-u_3$ the 
lengths of the two intervals and $u_3-u_2$ their distance) the entanglement entropy is straightforwardly written down from 
Eq.~(\ref{Sngen}).
Specialising to the R\'enyi mutual information in Eq.~(\ref{renyi MI def}), we have   
\be
\label{MI n N2 cft t-dep}
I^{(n)}_{A_1,A_2}
=
\frac{\pi c(n+1)}{12\tau_0 n}\,
\big[
q(t,u_3-u_1) + q(t,u_4-u_2) - q(t,u_4-u_1) - q(t,u_3-u_2) 
\big]\,.
\ee
Notice that, once the explicit expressions for the $q$'s from (\ref{qdef}) have been inserted in (\ref{MI n N2 cft t-dep}), the linear combination within the square brackets is such that only the terms involving the max's remain.
For large $t$, we have that $I^{(n)}_{A_1:A_2}$  vanishes for all $n$.

Taking the limit $u_3 \rightarrow u_2$ in (\ref{MI n N2 cft t-dep}), we get $I^{(n)}$ for two adjacent intervals 
\be
\label{MI n N2 cft t-dep adj}
I^{(n)}_{A_1,A_2}=
\frac{\pi c(n+1)}{12\tau_0 n}\,
\big[\,
t+ q(t,u_2-u_1) + q(t,u_4-u_2) - q(t,u_4-u_1) 
\big]\,.
\ee
It is worth mentioning that the time evolution of the entanglement entropy and mutual information 
has been also considered in the framework of holographic approach to CFTs~\cite{qe-hol}.

\section{Entanglement negativity}
\label{sec ent neg}

In this section we present the original part of the CFT calculation of this manuscript concerning the temporal evolution of the negativity 
between two intervals after a global quench to a conformal Hamiltonian. 
We consider $A =A_1 \cup A_2$, where the intervals $A_1$ and $A_2$ can be either adjacent or disjoint, 
as in Fig.~\ref{fig line intervals}.
An important special case is when $A$ is the entire system (i.e. $B \to \emptyset$).

The quantum field theory approach to  the logarithmic negativity $\cal{E}$  is based on a replica trick \cite{us-letter,us-long}. 
Let us  consider the traces $\Tr (\rho_A^{T_2})^n$ of integer powers of $\rho_A^{T_2}$.
For $n$ even and odd, denoted by $n_e$ and $n_o$ respectively, we have
\bea
\Tr (\rho_A^{T_2})^{n_e}&=&
\sum_i \lambda_i^{n_e}= 
\sum_{\lambda_i>0} |\lambda_i|^{n_e}+ \sum_{\lambda_i<0} |\lambda_i|^{n_e}\,, 
\label{trne}\\
\Tr (\rho_A^{T_2})^{n_o}&=&
\sum_i \lambda_i^{n_o}= 
\sum_{\lambda_i>0} |\lambda_i|^{n_o}- \sum_{\lambda_i<0} |\lambda_i|^{n_o}\,,
\label{trno}
\eea
where $\lambda_i$ are the eigenvalues of $\rho_A^{T_2}$.
Clearly, the functional dependence of $\Tr (\rho_A^{T_2})^n$ on $|\lambda_i|$ depends on the parity of $n$. 
Setting $n_e=1$ in (\ref{trne}), we formally obtain $ \Tr |\rho_A^{T_2}|$, whose logarithm gives the logarithmic negativity $\mathcal{E}$.
Instead, if we set $n_o=1$ in (\ref{trno}), we just get the normalization $\Tr \rho_A^{T_2}=1$.
Thus, the analytic continuations from even and odd values of $n$ are different and
the trace norm that we are interested in is obtained
by performing the analytic continuation of the even sequence (\ref{trne}) at $n_e\to1$.
By introducing
\be
\label{neg renyis def}
\mathcal{E}^{(n)} \equiv 
\ln \Big[ \Tr\big(\,\rho_A^{T_{2}}\big)^n \Big]\,,
\ee
we have that $\mathcal{E}^{(1)} =0$ identically and the logarithmic negativity $\mathcal{E} $ is given by the following replica limit
\be
\label{neg replica limit}
\mathcal{E} = \lim_{n_e \rightarrow 1}  \mathcal{E}^{(n_e)} \,.
\ee

For future convenience we also introduce the ratios
\be
\label{R ratio def}
R_{n}
\equiv
\frac{ \Tr (\rho_A^{T_2})^n }{ \Tr \,\rho_A^n }
\quad \Longrightarrow \qquad
\ln (R_{n})
= \mathcal{E}^{(n)} +(n-1) S^{(n)}_{A_1 \cup A_2}  \quad {\rm and}\quad 
%
\mathcal{E} = \lim_{n_e \to 1} \ln (R_{n_e})\,.
\ee
This replica approach has been introduced in the context of CFT \cite{us-letter, us-long}, but later it has been applied and generalised to 
many other circumstances \cite{a-13,ctt-13,rr-14,kpp-14,c-13,lv-13,cabcl-14}.

In a 1+1 dimensional quantum field theory, the traces $\Tr(\rho_A^{T_{2}})^n$ can be computed through correlators of twist fields, 
as shown in Refs.~\cite{us-letter, us-long}. 
As we already reported in the previous section, $\Tr \rho_A^n$ for the union of $N$ disjoint intervals $A=\cup_{i=1}^N [u_{2i-1}, u_{2i}]$ 
is given by the correlator $\langle \prod_{i=1}^N \mathcal{T}_n (u_{2i-1}) \bar{\mathcal{T}}_n(u_{2i}) \rangle$.
Now let us take the partial transpose of the $j$-th interval $A_j$. The quantity $\Tr(\rho_A^{T_j})^n$ can be computed from the correlator above where the twist fields $\mathcal{T}_n$ and $\bar{\mathcal{T}}_n$ at the endpoints of $A_j$ are exchanged while the remaining ones stay untouched, i.e. $\Tr(\rho_A^{T_j})^n=\langle \dots \bar{\mathcal{T}}_n (u_{2j-1}) \mathcal{T}_n(u_{2j})  \dots\rangle$. 
This procedure can be generalized straightforwardly to the case where the partial transposition involves two or more intervals. 
The configurations including adjacent intervals can be obtained as a limit of the previous one, where the distances between the proper intervals vanish. 
After this limit,  $\mathcal{T}^2_n$ or $\bar{\mathcal{T}}_n^2$ occur at the joining point between a partial transposed interval and the adjacent one that has not been partial transposed. 
Thus, the ratio (\ref{R ratio def}) can be computed through the corresponding correlators of twist fields.
For instance, when $A$ consists of two intervals $A_1$ and $A_2$, 
if they are disjoint we need to consider 
$\Tr(\rho_A^{T_{2}})^n = \langle  \mathcal{T}_n (u_1) \bar{\mathcal{T}}_n(u_2)  \bar{\mathcal{T}}_n (u_3) \mathcal{T}_n(u_4)  \rangle$ 
while, when they are adjacent, $\Tr(\rho_A^{T_{2}})^n = \langle  \mathcal{T}_n (u_1) \bar{\mathcal{T}}^2_n(u_2)  \mathcal{T}_n(u_3) \rangle$.

Specialising to the case of a 1+1 dimensional CFT, in order to compute correlation functions of twist fields, we need the
scaling dimension of $\mathcal{T}^2_n$ and $\bar{\mathcal{T}}_n^2$ which depends on the parity of $n$ as \cite{us-letter, us-long}
\be
\label{delta2 def}
\Delta^{(2)}_n \equiv  
\left\{ \begin{array}{ll}
\displaystyle
\Delta_{n}
\hspace{.5cm}& 
\textrm{odd $n$\,,}
\\
\rule{0pt}{.5cm}
\displaystyle
2\Delta_{n/2}
\hspace{.5cm}& 
\textrm{even $n$\,,}
\end{array}
\right.
\ee
where $\Delta_{n}$ has been defined in (\ref{Delta_n def}).

At this point we have all the needed ingredients to study the temporal evolution of the logarithmic negativity 
after a global quench. We can apply the method of Ref.~\cite{cc-05-quench} to the proper correlators on the strip, 
which involve both $\mathcal{T}_n$ ($\bar{\mathcal{T}}_n$) and $\mathcal{T}^2_n$ ($\bar{\mathcal{T}}^2_n$).
This means that we have to slightly generalize the setup described in the previous section by taking into account  
correlators  on the strip of fields which can have different dimensions.
Instead of presenting general formulas, we find more instructive to limit ourselves to discuss few cases of two intervals in 
which we are interested.

\subsection{Bipartite systems}
\label{sec neg bipart}

Although trivial, it is useful to first discuss the case in which $A=A_1\cup A_2$ is the entire system and we consider 
the partial transpose with respect to $A_2$. Since the time dependent state $| \psi(t) \rangle$ is pure at any time, 
$\rho_A$ corresponds to a pure state, and we can use the standard results \cite{vw-01} that for a pure state
the logarithmic negativity is the R\'enyi entropy with $n=1/2$, i.e.  
\be
\label{neg equal renyi12}
\mathcal{E} (t) = S^{(1/2)}_{A_2} (t)\,,
\ee
independently of the Hamiltonian governing the time evolution. 

When the evolution is conformal, we can re-obtain this trivial result by using the path integral approach discussed in the previous section.
We need to evaluate $\langle\, \mathcal{T}^2_n (w_1) \bar{\mathcal{T}}^2_n(w_2) \rangle_{\rm strip}$ and then to analytically continue 
to real time. The strip two-point function is related to the one in the UHP which has the standard form 
\be
\label{twist squared uhp N=1}
\langle  \mathcal{T}^2_n (z_1) 
\bar{\mathcal{T}}^2_n(z_2) \rangle_{\rm UHP} 
=
\frac{c^{(2)}_{n}}{|(z_1 - \bar{z}_1)(z_2 - \bar{z}_2) \,\eta_{1,2}|^{\Delta^{(2)}_n}} {\cal F}(\eta_{1,2})\,,
\ee
where the constants $c^{(2)}_{n}$ are related to $c_n$ in a known way \cite{us-long}, but their value is not important for what follows.  
Transforming the UHP to the strip, we find
\be
\label{neg traces cft N=1}
\Tr (\rho_A^{T_2})^n
 =
\langle\, \mathcal{T}^2_n (w_1) 
\bar{\mathcal{T}}^2_n(w_2) \,\rangle_{\rm strip} 
=
c^{(2)}_{n}
\bigg(\frac{\pi}{2\tau_0} \bigg)^{2\Delta_n^{(2)}}
\, \Bigg|
\left(\, \prod_{a=1}^{2} \frac{z_a}{z_a - \bar{z}_a} \right)
\frac{1}{\eta_{1,2}}\,\Bigg|^{\Delta_n^{(2)}} {\cal F}(\eta_{1,2})  \, ,
\ee
where $z_a = e^{\pi (u_a+\textrm{i}\tau) / (2\tau_0)}$.

The time evolution of the powers of the partial transpose comes from the analytic continuation in Eq.~(\ref{analytic cont}).
As usual, in the space-time scaling limit ($t \gg \tau_0$ and $u_2-u_1 \gg \tau_0$), we should retain only the leading behaviour  
of the above expression and all the various constants and the function ${\cal F}$ can be dropped, obtaining 
\be
\label{renyi neg N2 cft t-dep}
\mathcal{E}^{(n)} =
-\frac{\pi \Delta_n^{(2)}}{\tau_0}\,
\big[t + q(t, u_2-u_1)\big]\,, \qquad \Rightarrow \quad
\mathcal{E}=
\frac{\pi c}{4\tau_0}
\big[t + q(t, u_2-u_1)\big]\,,
\ee
which coincides with the R\'enyi entropy for $n=1/2$, as it should from (\ref{neg equal renyi12}).

We need to comment at this point on the asymptotic large time value of the negativity. 
Indeed, it is obvious that the negativity of one interval with respect to the rest of the system {\it does not thermalise}, 
being very different from the finite temperature negativity calculated in \cite{us-neg-T} (see also \cite{aw-08,fcga-08}). 
This is not a surprise since, by construction, this negativity is not a local quantity because it requires 
the partial transposition with respect to the infinitely large part $A_2$.

\subsection{Two adjacent intervals}
\label{sec neg two adjacent}

The conformal evolution of the entanglement negativity between two adjacent intervals after a global quench can be studied 
by considering the three point function $\langle \mathcal{T}_n \bar{\mathcal{T}}^2_n  \mathcal{T}_n \rangle$ on the strip which can be 
obtained by the mapping from the UHP of the three-point function 
\be
\label{two attached alphagamma}
\langle 
\mathcal{T}_n(z_1) \bar{\mathcal{T}}^2_n(z_2) \mathcal{T}_n(z_3) 
\rangle_{\rm UHP} 
\,=\,
 \frac{c_n}{\prod_{a=1}^{3} | z_a - \bar{z}_a |^{\Delta_{(a)}}}
\left(
\frac{\eta_{1,3}^{\Delta_n^{(2)}-2\Delta_n}}{\eta_{1,2}^{\Delta_n^{(2)}}  \eta_{2,3}^{\Delta_n^{(2)}}}
\right)^{1/2} {\cal F}(\{\eta_{j,k}\}) \, ,
\ee
where $\Delta_{(1)}=\Delta_{(3)}=\Delta_n$ and $\Delta_{(2)}=\Delta^{(2)}_n$ (given by (\ref{Delta_n def}) and (\ref{delta2 def}) respectively),
the harmonic ratios $\eta_{j,k}$ are defined in Eq.~(\ref{eta cross ratio uhp}) and again the function ${\cal F}$ depends 
on the full operator content of the theory and it is very difficult to calculate (see Refs.~\cite{us-long,ctt-14} for some explicit examples).
However, as it should be already clear at this point, we do not need this function in the space-time scaling limit, but we only need 
to ensure that the powers in the rest of the expression have been chosen in such a way that ${\cal F}$ is constant in the limits 
$\eta_{j,k}\to 0$ or $\eta_{j,k}\to 1$.
This can be easily checked by  employing the following OPE 
\be
\label{ope TT2}
\mathcal{T}_n(z) \bar{\mathcal{T}}^2_n(w) 
=
\frac{
C_{\mathcal{T}_n \bar{\mathcal{T}}_n^2\bar{\mathcal{T}}_n}
}{
|z-w|^{\Delta_n^{(2)}}} \, 
\bar{\mathcal{T}}_n(z) 
+ \dots\,,
\qquad
w \rightarrow z\,.
\ee
Taking separately the limits $z_2 \to z_1$ and $z_2 \to z_3$ in (\ref{two attached alphagamma}), and using (\ref{ope TT2}) 
it should be clear that ${\cal F}$ is constant in both the interesting limits. 
Then, the three-point function on the strip can be straightforwardly written by a conformal mapping, obtaining 
\be
\label{3points strip}
\langle 
\mathcal{T}_n(u_1) \bar{\mathcal{T}}^2_n(u_2) \mathcal{T}_n(u_3) 
\rangle_{\rm strip} 
\,=\,
 c_n
\bigg(\frac{\pi}{2\tau_0} \bigg)^{\Delta}
  \prod_{a=1}^{3} \bigg| \frac{z_a}{z_a - \bar{z}_a} \bigg|^{\Delta_{(a)}}
\left(
\frac{\eta_{1,3}^{\Delta_n^{(2)}-2\Delta_n}}{\eta_{1,2}^{\Delta_n^{(2)}}  \eta_{2,3}^{\Delta_n^{(2)}}}
\right)^{1/2} {\cal F}(\{\eta_{j,k}\}) \, ,
\ee
where $\Delta = 2\Delta_n+\Delta^{(2)}_n$ and $z_a = e^{\pi (u_a+\textrm{i}\tau) / (2\tau_0)}$. 
The time evolution of $\Tr (\rho_A^{T_2})^n $ for two adjacent intervals is then obtained by the analytic continuation of the above to
$\tau=\tau_0 +\textrm{i} t$. 

The CFT prediction for the temporal dependence of $\mathcal{E}^{(n)}$ in the space-time scaling regime  
($t \gg \tau_0$ and $|u_j-u_i| \gg \tau_0$) is then found, as usual, by dropping the various multiplicative constants and the 
function ${\cal F}$, obtaining
\be
\label{En adj cft}
\mathcal{E}^{(n)}
=
-\frac{\pi}{2\tau_0}
\Big[
\big(2\Delta_n+\Delta_n^{(2)}\big) t
+ \Delta_n^{(2)} \big( q(t, u_2-u_1) + q(t, u_3-u_2)\big)
- (\Delta_n^{(2)}-2\Delta_n) \,q(t, u_3-u_1)
\Big] \, .
\ee
Notice that, since $\Delta_2^{(2)}=0$, for $n=2$ the terms containing $u_2$ do not contribute and therefore the curve $\mathcal{E}^{(2)}(t)$ displays a change in its slope only at $t=(\ell_1+\ell_2)/2$.
This is a consequence of the trivial fact that $\bar{\mathcal{T}}^2_2=\mathbb{I}$ in (\ref{3points strip}).
Finally, taking the replica limit of (\ref{En adj cft}), we find the evolution of the logarithmic negativity for adjacent intervals
\be
\label{neg 2adj t-dep cft}
\mathcal{E}
\,=\,
\frac{\pi c}{8\tau_0}\,
\big[
\, t  - q(t,u_3-u_1) + q(t,u_2-u_1) + q(t,u_3-u_2) 
 \big]\,.
\ee

Considering the ratio (\ref{R ratio def}) for two adjacent intervals, since $A=A_1 \cup A_2$ is the interval $[u_1, u_3]$ in the spatial slice of the strip, $\Tr \,\rho_A^n $ corresponds to $[u_1, u_3]$ and  (\ref{R ratio def}), in terms of the CFT quantities, reads
\be
\label{Radj ratio cft}
R_{n}
\equiv
\frac{
\langle \mathcal{T}_n(u_1) \bar{\mathcal{T}}^2_n(u_2) \mathcal{T}_n(u_3) \rangle_{\rm strip} 
}{
\langle \mathcal{T}_n(u_1) \bar{\mathcal{T}}_n(u_3)\rangle_{\rm strip} 
}\,.
\ee
Notice that since $ \mathcal{T}^2_2= \bar{\mathcal{T}}^2_2=\mathbb{I}$, we have that $R_{2}=1$ identically.
The time dependence of $R_n$ in the space-time scaling regime is readily obtained by combining the expressions 
for the numerator and the denominator in (\ref{Radj ratio cft}),  finding
\be
\label{logA2t}
\ln ( R_{n})
\,=\,
\frac{\pi \Delta^{(2)}_n}{2\tau_0}
\big[
- t  + q(t,u_3-u_1) - q(t,u_2-u_1) -q(t,u_3-u_2) 
 \big]\,,
\ee
where, once the expressions for the $q$'s from (\ref{qdef}) have been plugged in, only the terms with the max's remain within 
the square brackets. 
This expression shows why the quantities $R_n$ are very useful when comparing these predictions with numerical calculations, 
indeed when comparing with Eq.~(\ref{En adj cft}) for ${\cal E}^{(n)}$ one immediately notices that all the dependence 
on $\Delta_n$ is not there.


%

\subsection{Two disjoint intervals}
\label{sec neg two disjoint}

The time evolution of logarithmic negativity between two disjoint intervals after a global quench can be computed 
from the analytic continuation of the four-point function 
$\langle \mathcal{T}_n \bar{\mathcal{T}}_n \bar{\mathcal{T}}_n \mathcal{T}_n \rangle$ on the strip 
(notice the order of the operators along the line which is crucial). 
The strip four-point function is derived from the conformal map from the same function on the UHP, 
which can be written as 
\be
\label{two disjoint T2 ansatz}
\langle \mathcal{T}_n(z_1) \bar{\mathcal{T}}_n(z_2) 
\bar{\mathcal{T}}_n(z_3)  \mathcal{T}_n(z_4) \rangle_{\rm UHP} 
\,=\,
\frac{c_n^2}{\prod_{a=1}^{4} |z_a - \bar{z}_a|^{\Delta_n}} 
\frac{1}{\eta_{1,2}^{\Delta_n}\,\eta_{3,4}^{\Delta_n}}
\left(
\frac{\eta_{1,4}\,\eta_{2,3}}{\eta_{1,3}\,\eta_{2,4}}
\right)^{\Delta^{(2)}_n/2-\Delta_n} {\cal F}(\{\eta_{j,k}\}) \, .
\ee
Again for the time evolution we do not need the knowledge of the function ${\cal F}$, but only to ensure that for 
$\eta_{j,k}\to 0$ and $\eta_{j,k}\to 1$ the form used above gives that ${\cal F}$ is constant. 
This can be easily checked by requiring that when the two intervals are far apart (i.e. $|z_3 - z_2| \to \infty$),
the four-point function factorizes into the product of two two-point functions and that for 
$z_3 \to z_2$ the following OPE holds
\be
\label{ope TT}
\bar{\mathcal{T}}_n(z) \bar{\mathcal{T}}_n(w) 
=
\frac{C^{\bar{\mathcal{T}}^2}_{\bar{\mathcal{T}} \bar{\mathcal{T}}}}{
|z-w|^{2\Delta_n - \Delta_n^{(2)}}} 
\, \bar{\mathcal{T}}^2_n(z)+ \dots\,,
\qquad
w \rightarrow z\,.
\ee
Mapping this four point-function on the strip, we get
\be
\label{two disjoint T2 ris}
\Tr (\rho_A^{T_2})^n 
\,=\,
c_n^2
\bigg(\frac{\pi}{2\tau_0} \bigg)^{\Delta}
 \prod_{a=1}^{4} \bigg| \frac{z_a}{z_a - \bar{z}_a} \bigg|^{\Delta_n}
\frac{1}{\eta_{1,2}^{\Delta_n}\,\eta_{3,4}^{\Delta_n}}
\left(
\frac{\eta_{1,4}\,\eta_{2,3}}{\eta_{1,3}\,\eta_{2,4}}
\right)^{\Delta^{(2)}_n/2-\Delta_n} {\cal F}(\{\eta_{j,k}\}) \, ,
\ee
being $\Delta = 4\Delta_n$ and $z_a = e^{\pi (u_a+\textrm{i}\tau) / (2\tau_0)}$.

%

The time evolution of $\mathcal{E}^{(n)}$ in the space-time scaling regime 
($t \gg \tau_0$ and $|u_j-u_i| \gg \tau_0$) is found by employing the analytic continuation (\ref{analytic cont}).
The result reads
\bea\fl
\label{En dis t-dep cft}
& & \hspace{-2.5cm}
\mathcal{E}^{(n)} =
-\frac{\pi}{\tau_0}
\bigg[\,
2\Delta_n\,t
+ \Delta_n \Big( q(t, u_2-u_1) + q(t, u_3-u_4)\Big) \\
& & \hspace{-.3cm}
- \big(\Delta_n^{(2)}/2 - \Delta_n\big)
\Big( q(t, u_4-u_1) + q(t, u_3-u_2) - q(t, u_3-u_1) - q(t, u_4-u_2) \Big)
\bigg] ,
\nonumber
\eea
whose replica limit is
\be
\label{logneg N2disj cft t-dep}
\mathcal{E}
\,=\,
\frac{\pi c}{8\tau_0}
\big[
q(t,u_3-u_1) + q(t,u_4-u_2) - q(t,u_4-u_1) - q(t,u_3-u_2) 
\big]\,.
\ee
The resulting expression for the negativity is identical to Eq.~(\ref{MI n N2 cft t-dep}) for the R\'enyi mutual information 
apart from the prefactor. 
Notice that in $\mathcal{E}^{(n)}$, for small $t$, the expression (\ref{En dis t-dep cft}) displays a linear regime in the initial 
part of the evolution, whose slope is $-2\pi\Delta_n/\tau_0$.
Being $\Delta_{n=1} = 0$, this linear regime does not occur for the logarithmic negativity (\ref{neg replica limit}).
We find it useful to compare (\ref{En dis t-dep cft}) with the corresponding quantity (\ref{En adj cft}) for adjacent intervals 
and observe that for $n=2$ all the dependences on $t$ and $u_j$ remain when the intervals are disjoint, while for adjacent 
intervals the $n=2$ is characterised by the cancellation of some variables.

The ratio $R_n$ in Eq.~(\ref{R ratio def}) for two disjoint intervals on the strip has the easy form 
\be
\label{two disjoint T2 ratio}
R_{n}
=
\frac{
\langle \mathcal{T}_n(u_1) \bar{\mathcal{T}}_n(u_2) 
\bar{\mathcal{T}}_n(u_3)  \mathcal{T}_n(u_4) \rangle_{\rm strip} 
}{
\langle \mathcal{T}_n(u_1) \bar{\mathcal{T}}_n(u_2) 
\mathcal{T}_n(u_3)  \bar{\mathcal{T}}_n(u_4) \rangle_{\rm strip} 
}
\,\simeq\,
\left(\frac{\eta_{1,4}\,\eta_{2,3}}{\eta_{1,3}\,\eta_{2,4}}\right)^{\Delta^{(2)}_n/2},
\ee
where we dropped the two functions ${\cal F}$ in numerator and denominator since they do not contribute in the
the space-time scaling limit after analytic continuation in time.  
From this expression, we get $R_{2}=1$ identically because $\mathcal{T}_2=\bar{\mathcal{T}}_2$.
Also the time evolution of this ratio in the space-time scaling regime is particularly easy:
\be
\label{renyi neg n N2 cft t-dep}
\ln R^{\rm}_{n}
\,=\,
\frac{\pi \Delta_n^{(2)}}{2\tau_0}\,
\big[
 q(t,u_4-u_1) + q(t,u_3-u_2) - q(t,u_3-u_1) - q(t,u_4-u_2) 
\big]\,.
\ee
Plugging the explicit expressions of the $q$'s (see (\ref{qdef})) in this expression, we find that only the terms involving the max's remain. 
Remarkably, also for disjoint intervals, these ratios $R_n$ do not depend on $\Delta_n$.

\begin{figure}[t]
\begin{center}
\includegraphics[width=0.8\textwidth]{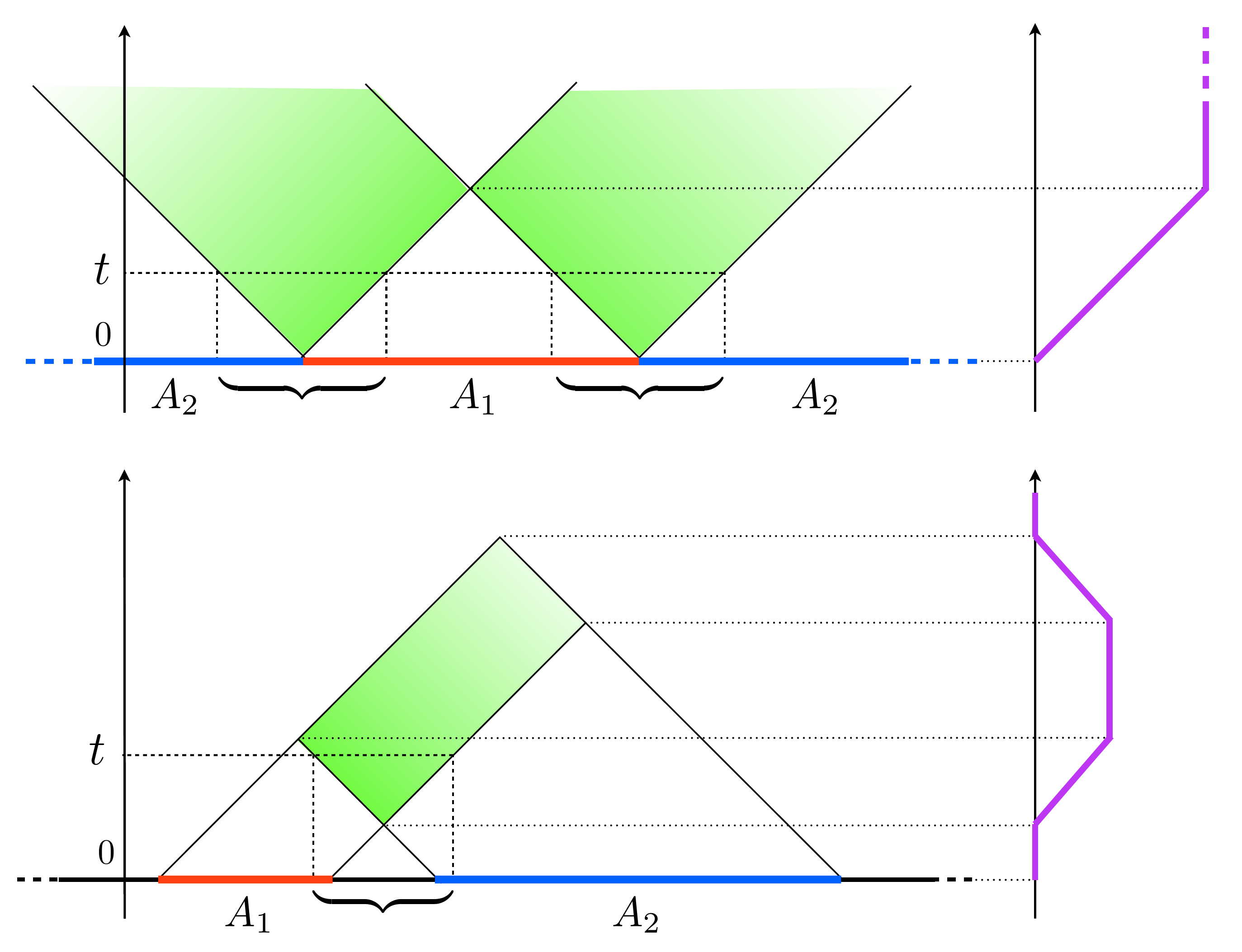}
\end{center}
\vspace{-.2cm}
\caption{
Graphical representation for the quasi-particle spreading of entanglement (for the case with 
all quasi-particles having the same velocity $v=1$ as in a CFT). 
The quasi-particles emitted from every point at $t=0$ and reaching one $A_1$ (red)
and the other $A_2$ (blue) are responsible of the entanglement between them. 
The entanglement at a given time $t$ is proportional to the section of the  
green shaded area, which is the intersection of the light cones starting from all the points of $A_1$ and $A_2$
(in the figure these lengths are the braces).
The time-dependence of the entanglement obtained in this way are depicted as purple curves on the right for a single interval 
in the infinite line (top) and two disjoint intervals (bottom): they are proportional to the CFT calculations in  
Eqs.~(\ref{SA one interval}) and (\ref{logneg N2disj cft t-dep}) respectively.
The regions from where the corresponding quasi-particles have been emitted at $t=0$ are obtained by projecting the intersections 
at time $t=0$ (vertical dashed lines).
}
\label{fig quasi-particles}
\end{figure}

\section{Quasi-particle interpretation and horizon effect}
\label{QP}

The time evolution of entanglement and total correlations after a quantum quench can be understood in terms of
the quasi-particles interpretation for the propagation of entanglement, first suggested in \cite{cc-05-quench}. 
According to this argument, since the initial state $|\psi_0\rangle$ has a very high energy relative to
the ground state of the Hamiltonian which governs the time evolution, it acts as a source of quasi-particle excitations.
Particles emitted from points further apart than the correlation length in the initial state are incoherent, but pairs of
particles emitted from a given point and subsequently moving to the left or right are highly
entangled and correlated. 
Let us suppose that a pair of quasi-particles with opposite momenta $(p,-p)$ is produced with a probability 
$\rho(p)$ (which depends on both the Hamiltonian governing the evolution and on the initial state).
After their production, these quasi-particles move ballistically with velocity $v_{p}=-v_{-p}$.
A quasi-particle of momentum $p$ produced at $x$ is therefore at $x+v_p t$ at time $t$.
In general there is also a maximum allowed speed of propagation $v_{\rm max}$ (which is 
connected with the existence of a Lieb-Robinson bound in a lattice model \cite{lr}). 

Now let us consider two regions of the system $A_1$ and $A_2$ (which can be either finite, infinite, semi-infinite, etc).
According to the argument in \cite{cc-05-quench},
the field at some point $x_1\in A_1$ will be entangled with that at a point $x_2\in A_2$ if a pair of entangled particles emitted from a
point $x$ arrive simultaneously at $x_1$ and $x_2$. 
The entanglement and the total correlation between $A_1$ and $A_2$ are proportional to the length of the interval
in $x$ for which this can be satisfied and it can be written as \cite{cc-05-quench}
\be
{\rm entanglement}\approx \int_{x_1\in A_1}dx_1\int_{x_2\in A_2}dx_2 
\int_{-\infty}^\infty dx\int dp \, \rho(p) f(p)  \,\delta(x_1-x-v_p t)\, \delta(x_2-x+v_p t) \, ,
\label{QPe}
\ee
where $f(p)$ is the contribution of the pair of quasi-particles to the given entanglement or correlation measure.

When all the quasi-particles move with the same speed $|v_p|=v$ 
(as in the case of a CFT discussed in the previous sections where we fixed $v=1$),
the $\delta$ functions do not depend on the momentum anymore and therefore the integral over $p$ gives just an overall normalisation (depending on the quantity we are considering), while the integral over the space coordinate can be easily done for arbitrary $A_1$ and $A_2$.
In particular, in the cases of one and two intervals, one straightforwardly recovers all the CFT expressions for the entanglement entropy, mutual information and negativity such as Eqs.~(\ref{Sngen}), (\ref{SA one interval}), (\ref{MI n N2 cft t-dep}) and (\ref{logneg N2disj cft t-dep}).
A graphical interpretation of this quasi-particle picture is reported in Fig.~\ref{fig quasi-particles}.

Furthermore, the above argument allows us also to understand what happens in the 
case of a non-linear dispersion relation leading to a mode dependent velocity, 
which will be fundamental for the interpretation and the understanding of the numerical 
data reported in the next section. 
Indeed, assuming that a maximum speed $v_{\rm max}$ exists, 
we have that the first linear increase of the entanglement is always present, but when $v_{\rm max}t$ equals
the half of some typical length of the configuration given by $A_1$ and $A_2$, the quasi-particles with velocity smaller than $v_{\rm max}$ start influencing the entanglement because we cannot ignore anymore
the integral over $p$ in Eq.~(\ref{QPe}). 
These slow quasi-particles lead to non-linear effects discussed e.g. for the entanglement entropy in Refs.~\cite{cc-05-quench,fc-08}. 
In particular we have that for very long times, the quasi-particles with approximately zero velocity govern the 
approach to the asymptotic value of the entanglement which usually is power-law as can be easily seen expanding Eq.~(\ref{QPe})
close to the points where $v_p=0$.

It is important to stress at this point that, while it was already established \cite{cc-05-quench,lk-08,fc-10} that the mutual information is 
correctly described by this quasi-particle picture, it is far from obvious that the same reasoning carries over 
to a complicated measure of the entanglement such as the negativity.
The previous section represents a proof of this fact in the context of CFT, while the following one will confirm it also 
for the harmonic chain.


\section{Numerical evaluation of the negativity and mutual information for the harmonic chain}
\label{app hc}

In this section we report the numerical evaluation of the time evolution of entanglement negativity and mutual information after a global quantum quench of the frequency parameter. 
The Hamiltonian of the periodic harmonic chain with nearest neighbour interactions is 
\be
\label{HC hamiltonian}
H(\omega) = \sum_{s=0}^{L-1} \left(
\frac{1}{2m}\,p_s^2+\frac{m\omega^2}{2}\,q_s^2 +\frac{K}{2}(q_{s+1} -q_s)^2
\right)\,,
\qquad  q_0 = q_L\,,
\qquad  p_0 = p_L\,.
\ee
where $L$ is the number of lattice sites of the chain, $m$ a mass scale, $\omega$ the one-particle oscillation frequency, 
and $K$ a nearest neighbour coupling.
The variables $p_i$ and $q_i$ satisfy standard commutation relations $[q_i,q_j] = [p_i,p_j] = 0$ and $[q_i,p_j] = \textrm{i}\delta_{ij}$.
We consider the harmonic chain because it is the only lattice model in which the partial transpose and the negativity can 
be obtained by means of correlation matrix techniques \cite{Audenaert02,br-04}. 
The model is critical  for $\omega=0$ and its continuum limit is conformal with central charge $c=1$. 
A canonical rescaling of the variables allows to rewrite it in a form where the parameters $\omega$, $m$, $K$ 
occur only in the global factor and in the coupling between nearest neighbour sites \cite{Audenaert02,br-04}.
We consider a quench in the parameter $\omega$, preparing the system in the ground state of (\ref{HC hamiltonian}) with 
$\omega=\omega_0\neq 0$ and letting the system evolve for $t>0$ with the critical Hamiltonian with $\omega=0$.
The quench dynamics of the Hamiltonian (\ref{HC hamiltonian}) has been studied already in several papers 
both on the lattice and in the continuum \cite{cc-07-quench,CE-08,bs-08,scc-09,pe-12,sc-14,rajapbour-14,dlsb-14}.

The Hamiltonian (\ref{HC hamiltonian}) is simply diagonalised in Fourier space in terms of standard  
annihilation and creation operators $a_k$ and $a_k^\dagger$. 
The diagonal form for the Hamiltonian is
\be
\label{HC hamiltonian diag}
H(\omega) = \sum_{k=0}^{L-1} \omega_k \left( a^\dagger_k a_k +\frac{1}{2} \right)\,,
\ee
where the dispersion relation is given by 
\be
\label{hc disp rel evolution}
\omega_k \equiv \sqrt{\omega^2+ \frac{4K}{m}  \sin^2(\pi k/L)}
\, \geqslant \omega\,,
\qquad k=0,\dots, L-1\,.
\ee
Notice that the Hamiltonian has a zero-mode for $k=0$ and $\omega=0$. 
This usually prevents a straightforward analysis of the critical behaviour, but for the global quench this will not be a problem, as we 
will see soon. 
From the dispersion relation, we straightforwardly have the velocity of each momentum mode as 
\be
v_k \equiv \frac{\partial \omega_k}{\partial p_k} = 
\, \frac{(K/m)\sin(p_k)}{\sqrt{\omega^2+ (4K/m)\sin^2(p_k/2)}}\,,
\qquad
p_k \equiv \frac{2\pi k}{L}\,,
\label{vk}
\ee
and the maximum one 
\be
\label{vmax def}
v_{\rm max} \equiv \textrm{max}_k (v_k)\,,
\ee
which determines the spreading of entanglement and correlations.
Notice that for $\omega=0$, $v_{\rm max}=1$ for $K=m=1$. 

In the quench protocol in which we are interested in, the system is prepared in the ground state $| \psi_0 \rangle$ of the Hamiltonian
\be
\label{HC hamiltonian diag 0}
H(\omega_0) = \sum_{k=0}^{L-1} \omega_{0,k} \left( a^\dagger_{0,k} a_{0,k} +\frac{1}{2} \right) ,
\ee
whose dispersion relation $\omega_{0,k}$ is Eq.~(\ref{hc disp rel evolution}) with $\omega=\omega_0$.
At $t=0$ the frequency parameter is suddenly quenched from $\omega_0$ to a different value $\omega$ 
and the system unitarily evolves through the new Hamiltonian (\ref{HC hamiltonian diag}), namely
\be
| \psi(t) \rangle =  e^{-\textrm{i} H(\omega) t} \, | \psi_0 \rangle\,,
\qquad
t>0\,.
\ee
In order to study the entanglement for the harmonic chain, we need to know the following two-point correlators
\be
\label{time dep corrs}
\begin{array}{l}
\mathbb{Q}_{r,s}(t) \equiv \langle \psi_0 | q_r(t)  q_s(t)  | \psi_0 \rangle \,,
\\
\rule{0pt}{.6cm}
\mathbb{P}_{r,s}(t) \equiv \langle \psi_0 | p_r(t)  p_s(t)  | \psi_0 \rangle \,,
\\
\rule{0pt}{.6cm}
\mathbb{M}_{r,s}(t) \equiv \langle \psi_0 | q_r(t)  p_s(t)  | \psi_0 \rangle \,,
\end{array}
\ee
where $q_r(t)$ and $p_r(t) $ are the time evolved operators in the Heisenberg picture
\be
q_r(t) = e^{\textrm{i} H t}  q_r(0) e^{-\textrm{i} H t} \,,
\qquad
p_r(t) = e^{\textrm{i} H t}  p_r(0) e^{-\textrm{i} H t} \,.
\ee
These correlators can be written as (see also \cite{rajapbour-14} for a slightly different approach)
\bea
\label{Qmat t-dep}
& & 
\mathbb{Q}_{r,s}(t)
\;=\;
\frac{1}{2L} \sum_{k=0}^{L-1} Q_k(t) \cos\Big[(r-s)\frac{2\pi k}{L}\Big] \,,
\\
 \label{Pmat t-dep}
& & 
\mathbb{P}_{r,s}(t)
\;=\;
\frac{1}{2L} \sum_{k=0}^{L-1} P_k(t)  \cos\Big[(r-s)\frac{2\pi k}{L}\Big]\,,
 \\
 \label{Mmat t-dep}
& & 
\mathbb{M}_{r,s}(t)
\;=\; \frac{\textrm{i}}{2}\, \delta_{r,s} -
\frac{1}{2L} \sum_{k=0}^{L-1} M_k(t)  \cos\Big[(r-s)\frac{2\pi k}{L}\Big] \,,
\eea
where  we collected the dependence on $\omega$, $\omega_0$ and $t$ into 
\bea
\label{Qmat t-dep-k}
& &  
Q_{k}(t)
\;\equiv\;  \frac{1}{m \omega_k}
\left( \,\frac{\omega_k}{\omega_{0,k}} \cos^2(\omega_k t) 
+ \frac{\omega_{0,k}}{\omega_k} \sin^2(\omega_k t) \right)\,,
\\
\rule{0pt}{.8cm}
 \label{Pmat t-dep-k}
& &
P_{k}(t)
\;\equiv\;  m \omega_k
\left( \,\frac{\omega_k}{\omega_{0,k}}  \sin^2(\omega_k t)
+ \frac{\omega_{0,k}}{\omega_k}  \cos^2(\omega_k t) \right)\,,
 \\
 \rule{0pt}{.8cm}
 \label{Mmat t-dep-k}
& & 
M_{k}(t)
\;\equiv\; 
\left(\,\frac{\omega_k}{\omega_{0,k}} - \frac{\omega_{0,k}}{\omega_k}  \right)
\sin(\omega_k t) \cos(\omega_k t) \,.
\eea
We notice that, for $t>0$ and $\omega =0$ the contribution from the mode $k=0$ is finite, indeed 
\be
Q_{0}(t) \,=\,  \frac{1}{m}
\left( \, \frac{1}{\omega_0} + \omega_0 t^2 \right)\,,
\qquad
P_{0}(t) \,=\, m \omega_0 \,,
\qquad
M_{0}(t) \,=\, -   \,\omega_0 t  \,.
\ee
The other modes are clearly always finite, and so we can consider 
the global quench to a massless Hamiltonian (while, as well known, we cannot set $\omega=0$ for the equilibrium properties). 


From the correlation functions, the entanglement entropy and negativity are constructed by standard methods. 
Indeed, given a subsystem $A$ of the lattice made by $\tilde{\ell}$ sites which could be either all in one interval or splitted 
in many disjoint intervals,
the reduced density matrix for $A$ can be studied by constructing the $\tilde{\ell} \times \tilde{\ell}$ matrices  
$\mathbb{Q}_A$, $\mathbb{P}_A$ and $\mathbb{M}_A$, which are the restrictions to the subsystem 
$A$  of the matrices $\mathbb{Q}$, $\mathbb{P}$ and $\mathbb{M}$ respectively \cite{pc-99, Audenaert02, br-04,  pedc-05}.
Given $\mathbb{Q}_A$, $\mathbb{P}_A$ and $\mathbb{M}_A$,  the covariance matrix 
$\gamma_A$ associated to the subsystem $A$ and the symplectic matrix $J_A$ of the corresponding size are
\be
\gamma_A(t) \equiv\, 
\textrm{Re}
\begin{pmatrix}
\mathbb{Q}_A(t) & \mathbb{M}_A(t)  \\
\mathbb{M}_A(t)^{\textrm{t}}  & \mathbb{P}_A(t) 
\end{pmatrix} ,
\qquad
J_A \equiv 
\begin{pmatrix}
\boldsymbol{0}_{\tilde{\ell}} & \mathbb{I}_{\tilde{\ell}} \\
- \mathbb{I}_{\tilde{\ell}} & \boldsymbol{0}_{\tilde{\ell}}
\end{pmatrix} ,
\ee
where $\mathbb{I}_{\tilde{\ell}} $ is the $\tilde{\ell} \times \tilde{\ell}$ identity matrix and $\boldsymbol{0}_{\tilde{\ell}} $ is the $\tilde{\ell} \times \tilde{\ell}$ matrix with vanishing elements.
We remark that the matrix $\mathbb{M}(t)$ has a non trivial real part for $t>0$.
At this point we compute the spectrum of $\textrm{i} J_A \cdot \gamma_A(t)$ which can be written as 
$\{ \pm \lambda_a(t) ;  a=1, \dots \tilde{\ell}\}$ with  $\lambda_a(t) >0 $. 
The time dependent R\'enyi entropies as function of the eigenvalues $\lambda_a(t)$ are finally written as
\be
\label{renyi entropies hc}
\textrm{Tr} \, \rho_A(t)^n = \prod_{a\,=\,1}^{\tilde{\ell}}
\left[
\bigg( \lambda_a(t)+\frac{1}{2} \bigg)^n - \bigg( \lambda_a(t)-\frac{1}{2} \bigg)^n\,
\right]^{-1},
\ee
and the entanglement entropy as
\be
\label{EE hc}
S_A(t) = \sum_{a\,=\,1}^{\tilde{\ell}} 
\left[
\bigg( \lambda_a(t)+\frac{1}{2} \bigg) \ln\bigg( \lambda_a(t)+\frac{1}{2} \bigg) 
- \bigg( \lambda_a(t) -\frac{1}{2} \bigg) \ln\bigg( \lambda_a(t)-\frac{1}{2} \bigg) 
\,\right] .
\ee

In order to compute the negativity, we denote by $A = A_1 \cup A_2$ a subregion of the harmonic chain and 
we consider the partial transpose with respect to $A_2$. 
In the covariance matrix $\gamma_A$, the net effect of the partial transposition 
is the inversion of the signs of the momenta corresponding to the sites belonging to $A_2$ \cite{Audenaert02}.
Thus, introducing the $\tilde{\ell} \times \tilde{\ell}$ diagonal matrix $\mathbb{R}_{A_2}$ 
which has $-1$ in correspondence of the sites of $A_2$ and $+1$ otherwise, we can construct 
\be
\gamma_A^{T_2}(t) 
\equiv\,
\begin{pmatrix}
 \mathbb{I}_{\tilde{\ell}} & \boldsymbol{0}_{\tilde{\ell}}  \\
  \boldsymbol{0}_{\tilde{\ell}}  & \mathbb{R}_{A_2}
\end{pmatrix}
\cdot 
\gamma_A(t) 
\cdot 
\begin{pmatrix}
 \mathbb{I}_{\tilde{\ell}} & \boldsymbol{0}_{\tilde{\ell}}  \\
  \boldsymbol{0}_{\tilde{\ell}}  & \mathbb{R}_{A_2}
\end{pmatrix} ,
\ee 
and compute the spectrum of $\textrm{i} J_A \cdot \gamma_A^{T_2}(t)$, 
which again can be written as $\{ \pm \chi_a(t) ;  a=1, \dots \tilde{\ell}\}$ with  $\chi_a(t) >0 $.
Then, the trace of the $n$-th power of $\rho_A^{T_2}$ is
\be
\label{renyi negativities hc}
\textrm{Tr} (\rho_A^{T_2})^n = \prod_{a\,=\,1}^{\tilde{\ell}}
\left[
\bigg( \chi_a(t)+\frac{1}{2} \bigg)^n - \bigg( \chi_a(t) -\frac{1}{2} \bigg)^n\,
\right]^{-1},
\ee
while the trace norm reads
\be
\label{trace norm hc}
|| \rho_A^{T_2} || 
 = 
\prod_{a\,=\,1}^{\tilde{\ell}}
\Bigg[\,
\bigg| \chi_a(t)+\frac{1}{2} \bigg| - \bigg| \chi_a(t) -\frac{1}{2} \bigg| \,
\Bigg]^{-1}
 = 
 \prod_{a\,=\,1}^{\tilde{\ell}}
 \textrm{max}
 \bigg[ 1 ,  \frac{1}{2\chi_a(t)} \bigg]\,,
\ee
which gives the logarithmic negativity $\mathcal{E}=\ln (|| \rho_A^{T_2} || )$.

In what follows we will compute entanglement entropies and negativity from the above formulas by calculating the 
spectrum of the appropriate covariance matrix. 
Since the parameters $K$ and $m$ can be absorbed in a redefinition of the canonical variables, we fix them to $K=m=1$ and we just consider a quench in the frequency (or mass) parameter from $\omega_0$ to $\omega$.
The data obtained by numerically diagonalising the covariance matrix  
for several bi- and tripartitions of the harmonic chain are reported in the following 
subsections.

\subsection{The entanglement entropy of one interval}

\begin{figure}[t]
\vspace{.2cm}
\includegraphics[width=\textwidth]{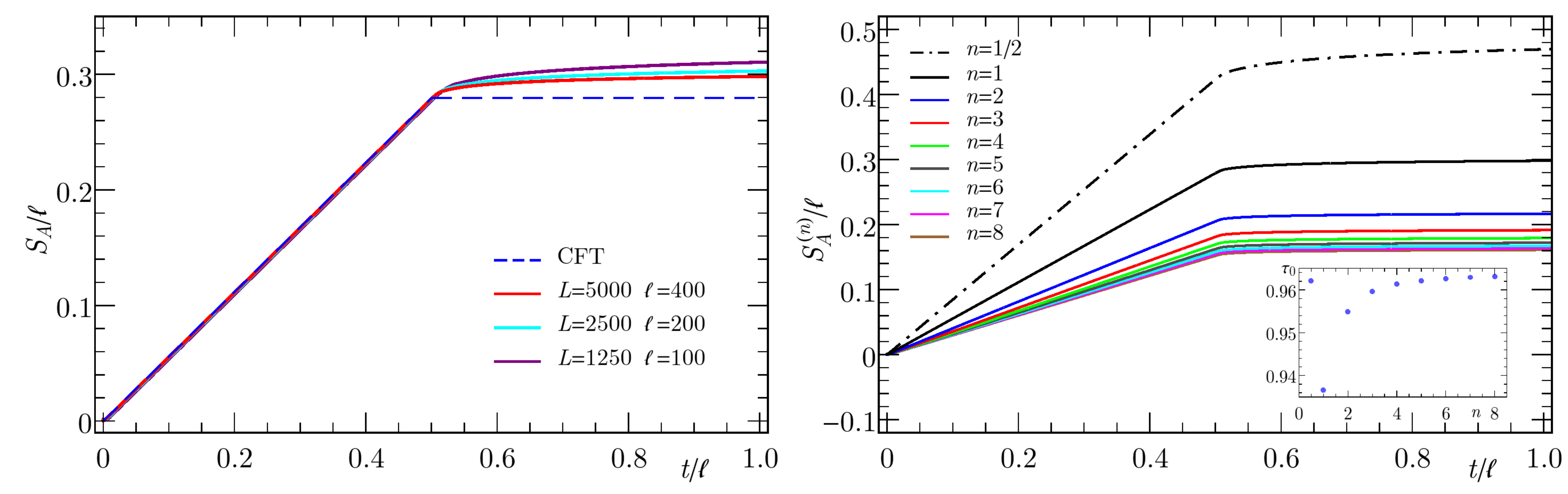}
\caption{Left:
Temporal evolution of the entanglement entropy for one interval of $\ell$ sites in a periodic harmonic chain with $L$ sites. At $t=0$ the mass is quenched from $\omega_0=1$ to $\omega=0$. The dashed curve is the CFT prediction (\ref{SA one interval}) with $c=1$ and the best fitted value for $\tau_0$.
Right: Temporal evolution of the R\'enyi entropies and of the logarithmic negativity $\mathcal{E}_A$, which coincides with $S_A^{(1/2)}$ in this case, for a periodic chain with $L=5000$ and $\ell=400$. In the inset, we report the best fitted values of $\tau_0$ 
for the values of $n$ displayed in the main plot.
}
\label{fig ee N1}
\end{figure}

It is instructive to start our analysis from the study of the quench dynamics of the entanglement entropy of 
a single interval, although this has been recently studied in Ref.~\cite{rajapbour-14}.
Indeed, this preliminary analysis allows us to understand the regime of applicability of the CFT 
and the optimal quench parameters in order to observe a CFT scaling.

In Fig.~\ref{fig ee N1}, we report the time evolution of the entanglement entropy for a quench from $\omega_0=1$ 
to $\omega=0$. 
We consider finite chains of length up to $L=5000$ and several values of $\ell\ll L$ with $\ell/L$ kept fixed.
It is evident from the figure that the behaviour of von Neumann and R\'enyi entropy is in good qualitative agreement with the 
CFT prediction (\ref{SA one interval}) with a linear growth for $t<\ell/2$ followed by saturation for $t>\ell/2$ 
(we recall that for $\omega=0$  the maximum mode velocity is $v_{\rm max}=1$, cf. Eq.~(\ref{vmax def})).
However, few comments on these results are needed. 
First, for $t>\ell/2$ the entanglement entropies do not saturate but they 
show a slow growth toward an asymptotic value. This is a well known phenomenon \cite{cc-05-quench,fc-08}
and it is due to the entanglement generated by slow quasi-particles moving with velocity $v_k<v_{\rm max}=1$
as in Eq.~(\ref{vk}). A second comment concerns the fitted value of $\tau_0$: this is shown in the inset of Fig.~\ref{fig ee N1} as function of the order of the R\'enyi entropy $n$. There is a minor dependence on $n$, as already anticipated and noticed in the 
literature  \cite{rajapbour-14}, but overall $\tau_0$ is very close to the initial correlation length 
$\xi_0\sim \omega_0^{-1}=1$.
Finally we must comment on the chosen value of $\omega_0=1$.
In preliminary calculations we considered several values of $\omega_0$ which however we do not report here,
but $\omega_0=1$ is the one for which the conformal scaling describes the data more accurately. 
This can be easily understood from the fact that (i) we should be in the regime $t,\ell\gg \omega_0^{-1}$ requiring that 
the initial frequency should not be too small,
(ii) we should be in a regime in which the continuum description is appropriate, requiring the initial correlation length not to be too small (i.e.\ $\omega_0$ not too large), in order to avoid a magnification of lattice effects. 
The value $\omega_0\sim 1$ appears to be the best compromise between these two effects.

The R\'enyi entropy with $n=1/2$ corresponds to the logarithmic negativity, but, as expected, does not display
any peculiar behaviour compared with the other values of $n$.

\subsection{Two adjacent intervals}

\begin{figure}[t]
\vspace{.2cm}
\hspace{-.4cm}
\includegraphics[width=1.0\textwidth]{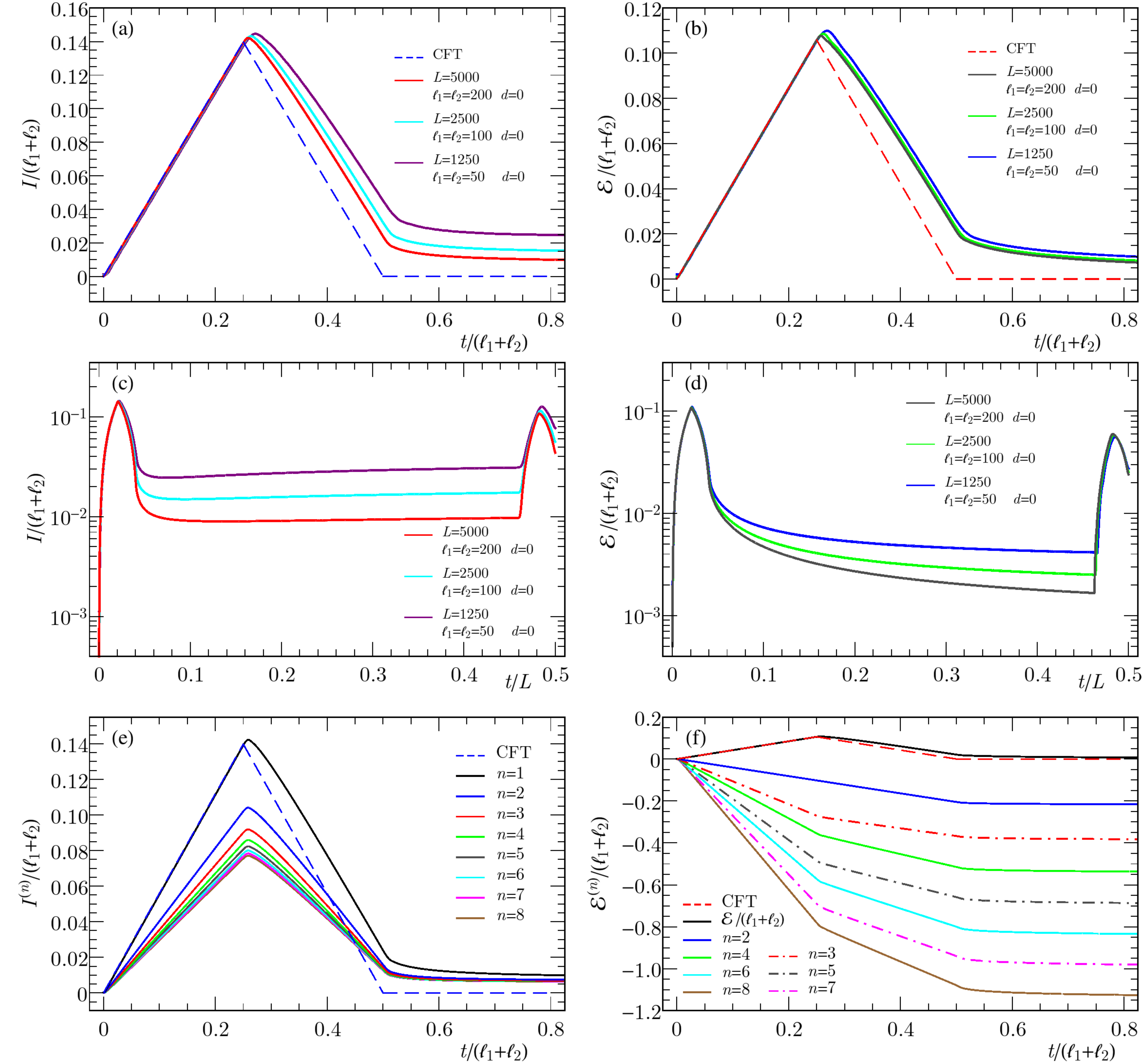}
\caption{
Adjacent intervals with several equal lengths $\ell_1=\ell_2$ and for various total size $L$ of the periodic harmonic chain. 
All panels show the data for $\omega_0=1$ and a critical evolution, $\omega=0$.
Panels (a) and (c) display the mutual information $I$ while (b) and (d) the logarithmic negativity $\mathcal{E}$.
Top and middle panels show different time scales and  the revivals due to the finiteness of the system are evident in 
the middle panels, where a larger range of $t$ is considered. 
The dashed CFT curves in (a) and (b) are given  by (\ref{MI n N2 cft t-dep adj}) and (\ref{neg 2adj t-dep cft}) respectively.
In the last two panels we show the time evolution of the R\'enyi mutual information $I^{(n)}$ (e) and of the replicated negativity  
$\mathcal{E}^{(n)}$ (f) for various values of $n$. 
}
\label{fig N2 L200L200d0}
\end{figure}

\begin{figure}[t]
\vspace{.2cm}
\hspace{-.4cm}
\includegraphics[width=1.0\textwidth]{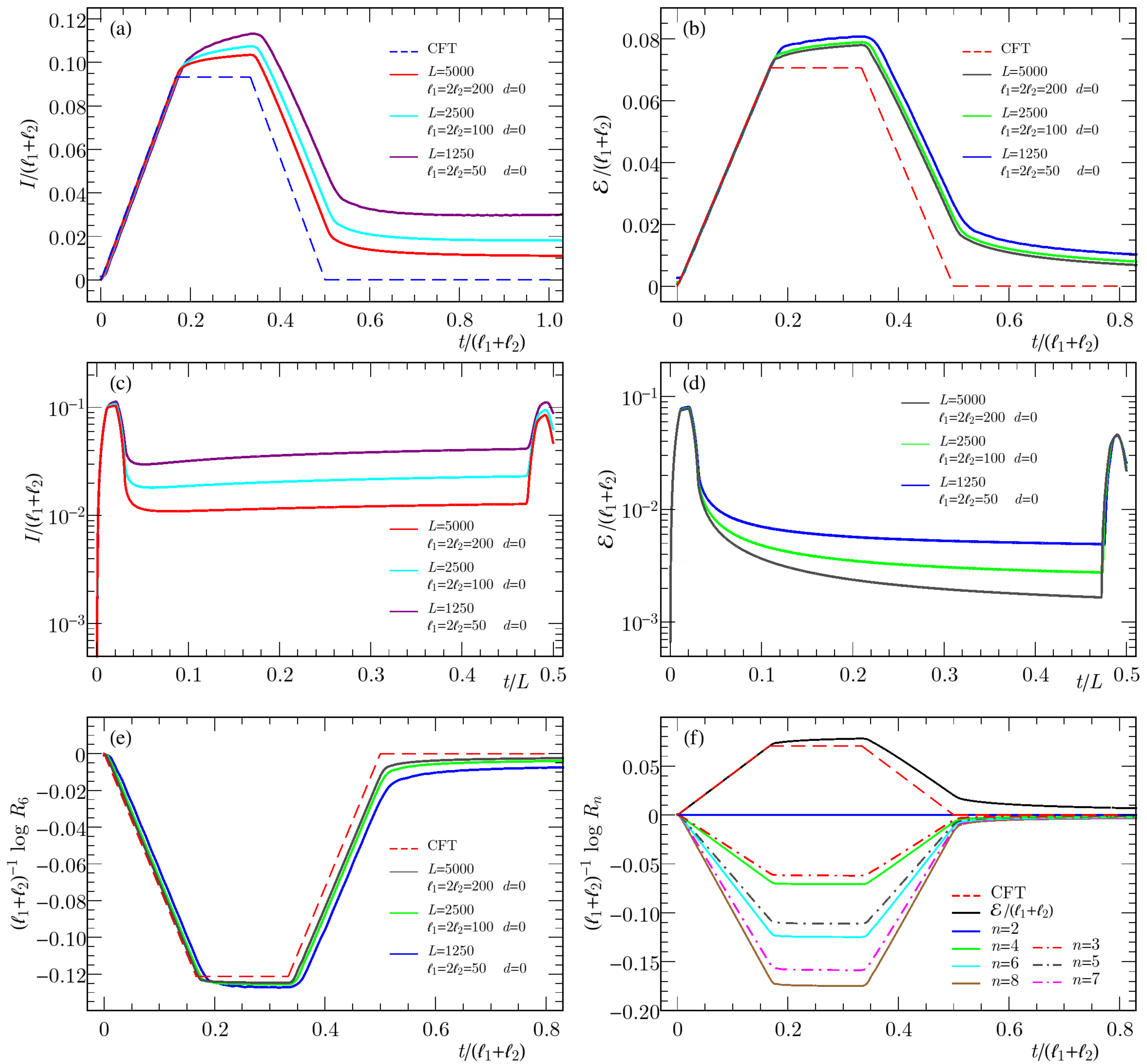}
\caption{
Adjacent intervals with different lengths $\ell_1=2\ell_2$ for different intervals lengths and total size $L$ of the periodic harmonic chain. Critical evolution of the mutual information $I$ (panel (a), (c)) and of the logarithmic negativity $\mathcal{E}$ (panels (b), (d)).
The revivals  are reported in panels (c) and (d). 
Notice that outside the light cone $\mathcal{E}$ always decays while $I$ reaches a plateau (apart from some
finite size effects responsible of a very slow increase) (panels (c) and (d)).
Compared to  Fig.~\ref{fig N2 L200L200d0} for the case of intervals with equal length, we observe 
a plateau starting at $t \simeq \textrm{min}(\ell_1,\ell_2)/2$ with temporal width $|\ell_2-\ell_1|/2$.
The panel (e) and (f) report the ratios $R_n$ for several values of $n$.
}
\label{fig N2 L200L100d0}
\end{figure}

We start the study of the entanglement between two different intervals from the case of adjacent ones. 
In the various figures that will follow we report both the mutual information and the negativity in order 
to simplify the discussion of similarity and differences. As already stressed above, 
the principal new results of this manuscript concern the evaluation of the negativity, since the time dependence of the 
mutual information in CFT was already established in \cite{cc-05-quench} and these results were checked in a few 
numerical works for the Ising chain \cite{fc-10} and the Bose-Hubbard model \cite{lk-08}, but, to the best of our knowledge, not 
for the harmonic chain. 

The numerical results for two adjacent intervals for a quench from $\omega_0=1$ to $\omega=0$
are shown in Figs.~\ref{fig N2 L200L200d0} and \ref{fig N2 L200L100d0}.
The main differences between the two sets of plots is that in the first one the two intervals have equal lengths, while in the second one the length of one interval is half of the other. 
Let us discuss these two sets of plots critically. 
In the top panels of Fig.~\ref{fig N2 L200L200d0}, we report the time evolution of the mutual 
information (a) and logarithmic negativity (b) on time scales of the order of $\ell$. 
They have a very similar behaviour characterised by an initial linear growth, followed by an almost linear dropping up to time 
$t=\ell_1$ when a slow power-law relaxation to the asymptotic vanishing value starts.
These results are in agreement with the expectation from CFT and their behaviour is simply understood in terms of the 
quasi-particle picture as already explained in Sec.~\ref{QP}. Also the differences between the linear CFT behaviour and the 
non-linear one of the actual data is easily understood in terms of the slow quasi-particles in analogy to the 
entanglement entropy of a single interval in the previous subsection. 
No particular difference is observed in this regime between negativity and mutual information. 
In the two middle panels of Fig.~\ref{fig N2 L200L200d0}, we again report the time evolution of the mutual 
information (c) and logarithmic negativity (d), but on time scales of the order of the system's size $L$. 
The main feature in this case is the presence of quantum revivals at time equal to $t \simeq (L-[\ell_1+\ell_2])/2$.
A first important difference between the mutual information and the negativity appears on these time-scales,
indeed while the former reaches a plateau in a large time-window, the latter decreases monotonically until the revival. 
Finally, in the last two panels we also report the R\'enyi mutual information (e) and the replicated negativity ${\cal E}^{(n)}$.
Their behaviour is again in qualitative agreement with the CFT predictions and the differences are easily understood in terms 
of slow quasi-particles. 
However, we mention that ${\cal E}^{(n)}$ with $n>1$ does not follow the same behaviour as the mutual information or negativity  
and indeed it is a monotonically decreasing function of time, changing the slope in the time intervals identified above (i.e. 
at $t=\ell_1/2$ and at $t=\ell_1$).
This is not a surprise since ${\cal E}^{(n)}$ is not a measure of entanglement and neither a quantifier of the correlations.
However, as already stressed, the piece-wise quasi-linear behaviour is compatible with the CFT prediction apart 
from the effect of the slow modes.

In Fig.~\ref{fig N2 L200L100d0}, we report the case of two adjacent intervals of different lengths $\ell_1=2\ell_2$.
The main difference compared to the case of equal lengths is the appearance of a plateau after the first linear increase for 
both the mutual information and the negativity. This is in perfect agreement with the CFT results which indeed 
predict (see Eqs.~(\ref{MI n N2 cft t-dep adj}) and (\ref{neg 2adj t-dep cft})) a linear increase up to $t= {\rm min}(\ell_1, \ell_2)/2$, followed by a plateau for  
${\rm min}(\ell_1, \ell_2)/2< t < {\rm max}(\ell_1, \ell_2)/2$, a linear decrease for ${\rm max}(\ell_1, \ell_2)/2 < t < (\ell_1 + \ell_2)/2$, 
and finally zero constantly. 
It is evident that the differences between CFT and actual data for the harmonic chain are due to slow quasi-particles
which give corrections for $t> {\rm min}(\ell_1, \ell_2)/2$. 
In the panel (c) and (d) of Fig.~\ref{fig N2 L200L100d0} we show the evolution of mutual information and negativity 
for a larger time window and, as before, the revivals are apparent at time $t=(L-\ell_1-\ell_2)/2$.
Finally in the last two panels (e) and (f) we report the behaviour of the universal ratio $R_n$ as a function of time. 
Again there are no particular differences with the CFT prediction except for those ones due to slow quasi-particles. 
%
From the last panel (f), one notices that $R_{2}=1$ identically. 
In the quantum field theory approach, this exceptional behaviour for $n=2$ is easily understood from the fact that 
$ \mathcal{T}^2_2= \bar{\mathcal{T}}^2_2=\mathbb{I}$ (cf. Sec.~\ref{sec neg two disjoint}) and it is true also on the lattice.  
Notice that $\ln (R_n)$ has a behaviour which closely resembles the one of the mutual information and the 
negativity, showing that, in this particular case, they are somehow measuring the amount of correlations or entanglement.
This is an important difference compared to the quantity ${\cal E}^{(n)}$ in Fig.~\ref{fig N2 L200L200d0}.

\begin{figure}[t]
\vspace{.2cm}
\hspace{-.4cm}
\includegraphics[width=1.0\textwidth]{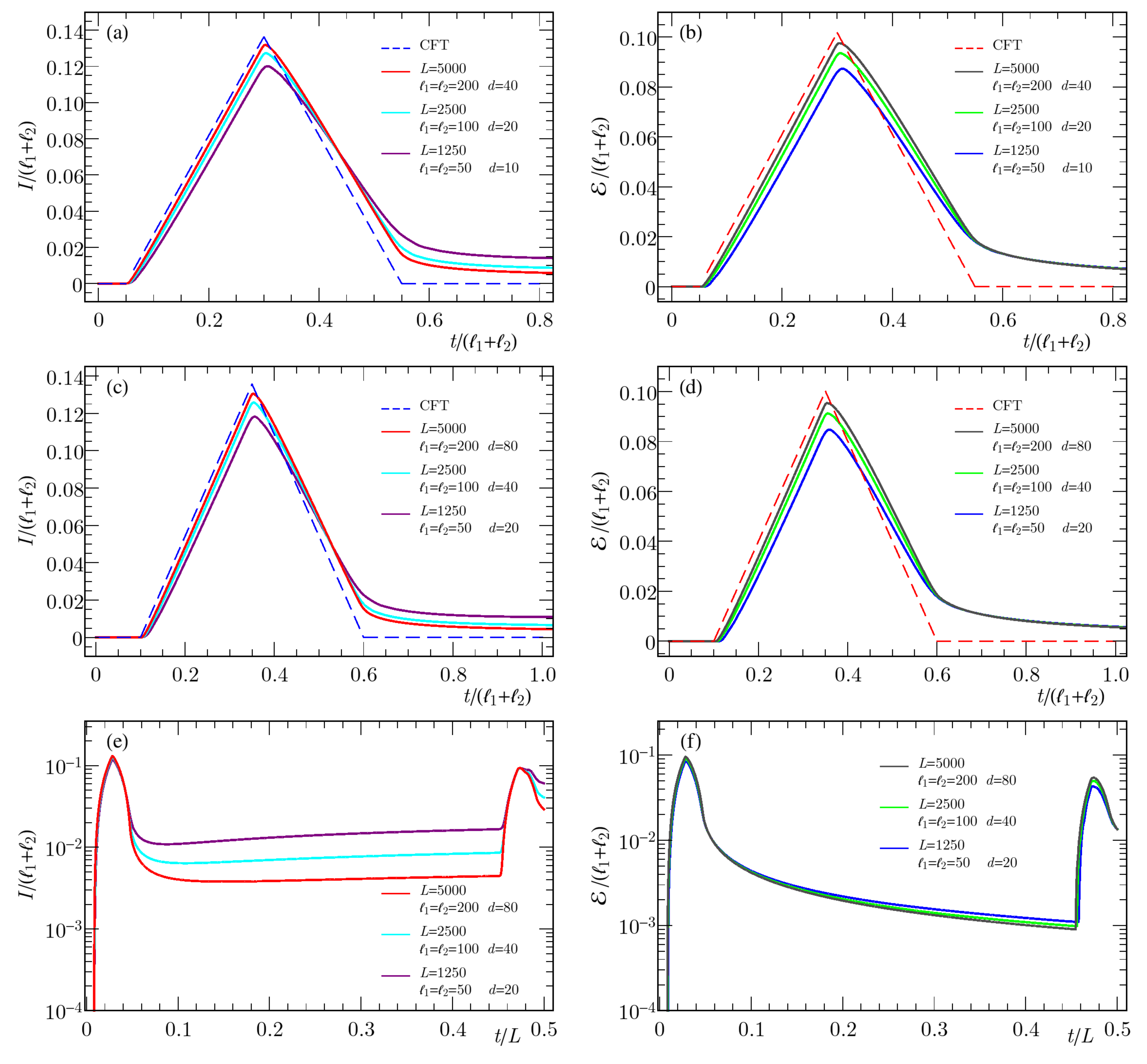}
\caption{
Disjoint intervals with equal lengths $\ell_1=\ell_2$ separated by $d$ sites, for various intervals lengths, separations $d$, 
and total size $L$ of the periodic harmonic chain. Time evolution of the mutual information $I$ ((a), (c) and (e)) and of the logarithmic 
negativity $\mathcal{E}$ ((b), (d), and (f)).
The revivals at large $t$ are evident in the bottom panels. 
}
\label{fig N2 L200L200d40}
\end{figure}


\subsection{Two disjoint intervals}

In this subsection we move to the case of two disjoint intervals. 
In Fig.~\ref{fig N2 L200L200d40} we report the numerically calculated mutual information and negativity 
for two disjoint intervals of equal length and for two different sets of distances between them. 
It is clear from the figure that the main difference compared to the case of adjacent interval is that there 
is an initial region for $t<d/2$ in which there is no entanglement and no correlations (we recall that the
initial correlation length is $\omega_0^{-1}=1$, so that the initial entanglement is very small). 
Then at time equal to $d/2$ the entanglement starts growing linearly, reaches a maximum and then decreases almost 
linearly. Again, this behaviour is compatible with the quasi-particle interpretation and also the power law 
relaxation for large times can be understood in terms of slow modes.  
Even in the case of disjoint intervals, we have studied the revivals.  The results, which are very similar to the ones 
shown in the case of adjacent intervals, are reported in Fig.~\ref{fig N2 L200L200d40}, panels (e) and (f).

\begin{figure}[t]
\vspace{.2cm}
\hspace{-.4cm}
\includegraphics[width=1.0\textwidth]{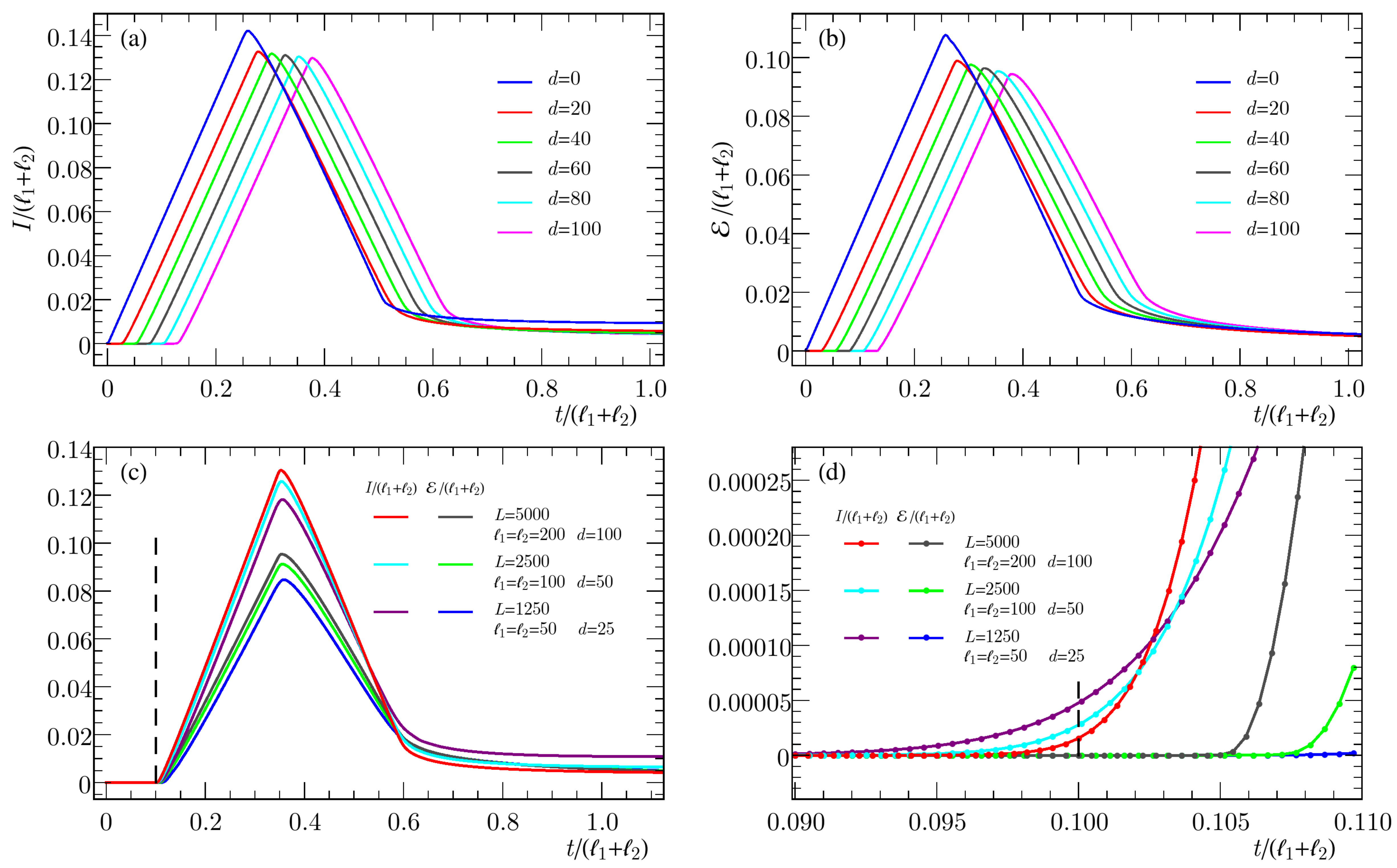}
\caption{
Mutual information (a) and the logarithmic negativity (b) of disjoint intervals with equal lengths $\ell_1=\ell_2=200$, separated by  
distances $d$  in a periodic chain with $L=5000$. 
(c) Comparison between $I$ and $\mathcal{E}$.
(d) Zoom of the initial growth: a late birth effect for $\mathcal{E}$ is observed, which decreases as the continuum limit is approached.
The vertical dashed line in (c) and (d) indicates the time $t=d/2$ where the growth should start in the continuum theory.
}
\label{fig N2 L200L200manyd}
\end{figure}

Conversely, a very interesting phenomenon can be observed by looking at very short times after the 
entanglement starts growing. This is illustrated in Fig.~\ref{fig N2 L200L200manyd}, where in the top panels 
we show the mutual information and the negativity as function of time for different distances between the two intervals. 
As before, the behaviour is very reminiscent of the CFT prediction. In the panel (c) of the same figure, 
we show in the same plot the negativity and the mutual information for fixed $d/\ell$ and looking closely 
to the time when the entanglement starts growing it is already clear that something is happening. 
For this reason in the panel (d) we zoom close to the point $t=d/2$ and we highlight a very peculiar phenomenon:
while the mutual information starts moving from zero slightly before $t=d/2$ 
the negativity starts slightly after $t=d/2$.
The behaviour of the mutual information is simply the exponential tails of the correlations outside the light-cone, 
but the behaviour of the negativity is new. 
From the figure, it is evident that increasing the total system size $L$  (at fixed ratios $d/L$ and $\ell/L$)
this phenomenon disappears in such a way to recover the CFT result and the quasi-particle interpretation of the evolution 
in the continuum limit. 
As a consequence, this remarkable phenomenon is a lattice effect and so 
cannot have an explanation in terms of quasi-particles, but it would be interesting to understand its precise origin. 
In a suggestive way, and in analogy with the famous 
sudden death of entanglement \cite{ye-09} (which will be discussed below), 
we term this phenomenon as the {\it late birth of entanglement}.


\begin{figure}[t]
\vspace{.2cm}
\hspace{-.4cm}
\includegraphics[width=1.0\textwidth]{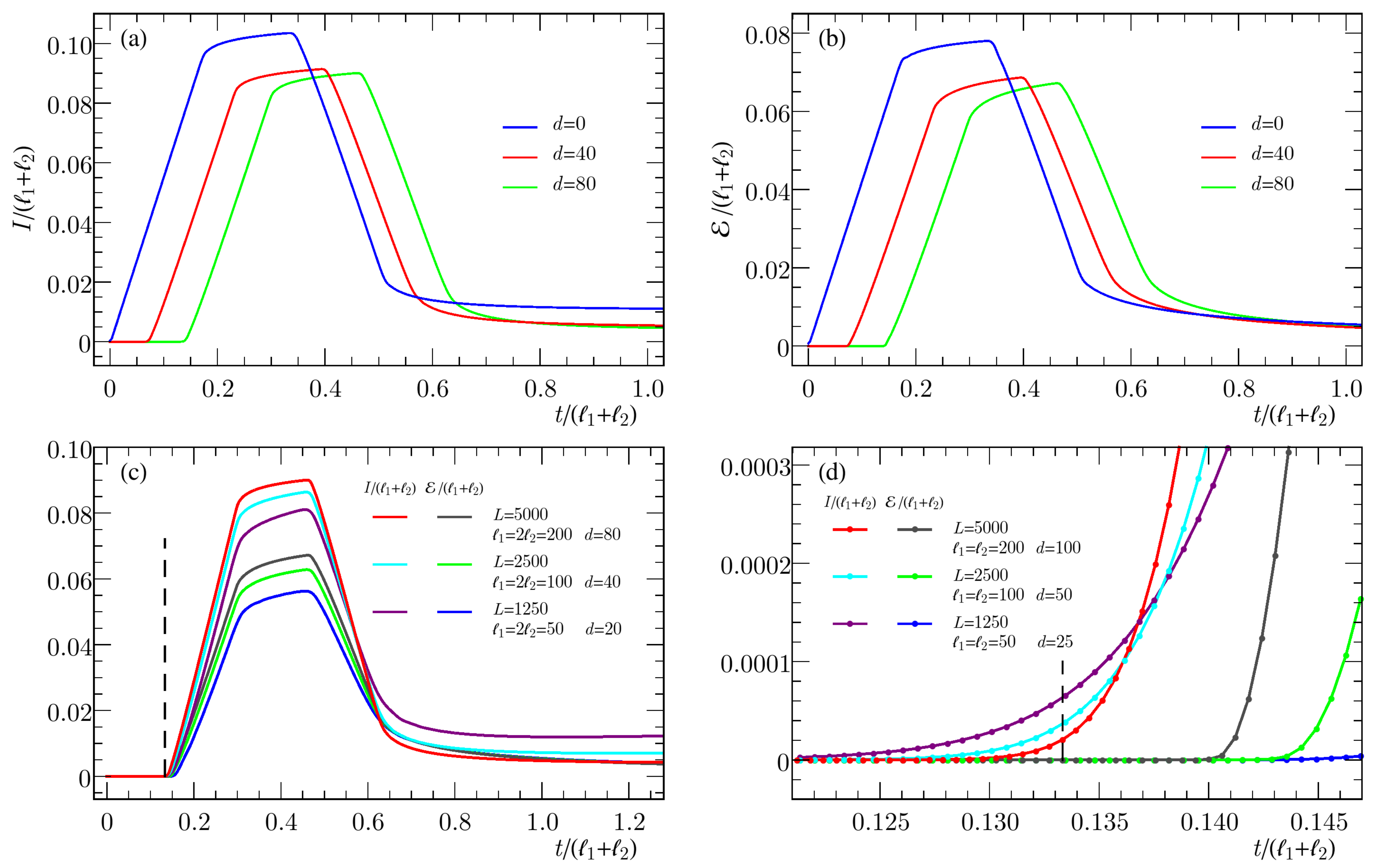}
\vspace{.1cm}
\caption{
Mutual information (a) and the logarithmic negativity (b) of disjoint intervals with different lengths $\ell_1=2\ell_2=200$, separated by various distances $d$ sites in a periodic chain with $L=5000$.
Comparison between $I$ and $\mathcal{E}$ in (c) and zoom on the initial growth in (d), to highlight the late birth of $\mathcal{E}$.
With respect to the corresponding plots in Fig.~\ref{fig N2 L200L200manyd}, here a plateau occurs whose temporal width is $|\ell_2-\ell_1|/2$. 
The vertical dashed line in (c) and (d) indicates the time $t=d/2$ where the growth should begin, according to the quasi-particles picture.
}
\label{fig N2 L200L100manyd}
\end{figure}


In Fig.~\ref{fig N2 L200L100manyd} we consider again the mutual information (a) and the negativity (b) for various distances between 
two intervals of different lengths $\ell_1=2\ell_2$.
The behaviour is in perfect agreement with the CFT prediction with a linear growth starting at $d/2$  up to $d/2+\textrm{min}(\ell_1, \ell_2)/2$, followed by a plateau lasting $|\ell_1-\ell_2|/2$, then a linear decrease up to $t= (d+\ell_1+\ell_2)/2$ and finally a power-law approach to zero. 
In the panel (c) we report on the same plot the negativity and the mutual information, observing
again the fingerprints of the late birth, which are straightforwardly confirmed by zooming for times close to $d/2$, as done in panel (d).

\begin{figure}[t]
\vspace{.3cm}
\hspace{-.4cm}
\includegraphics[width=1.\textwidth]{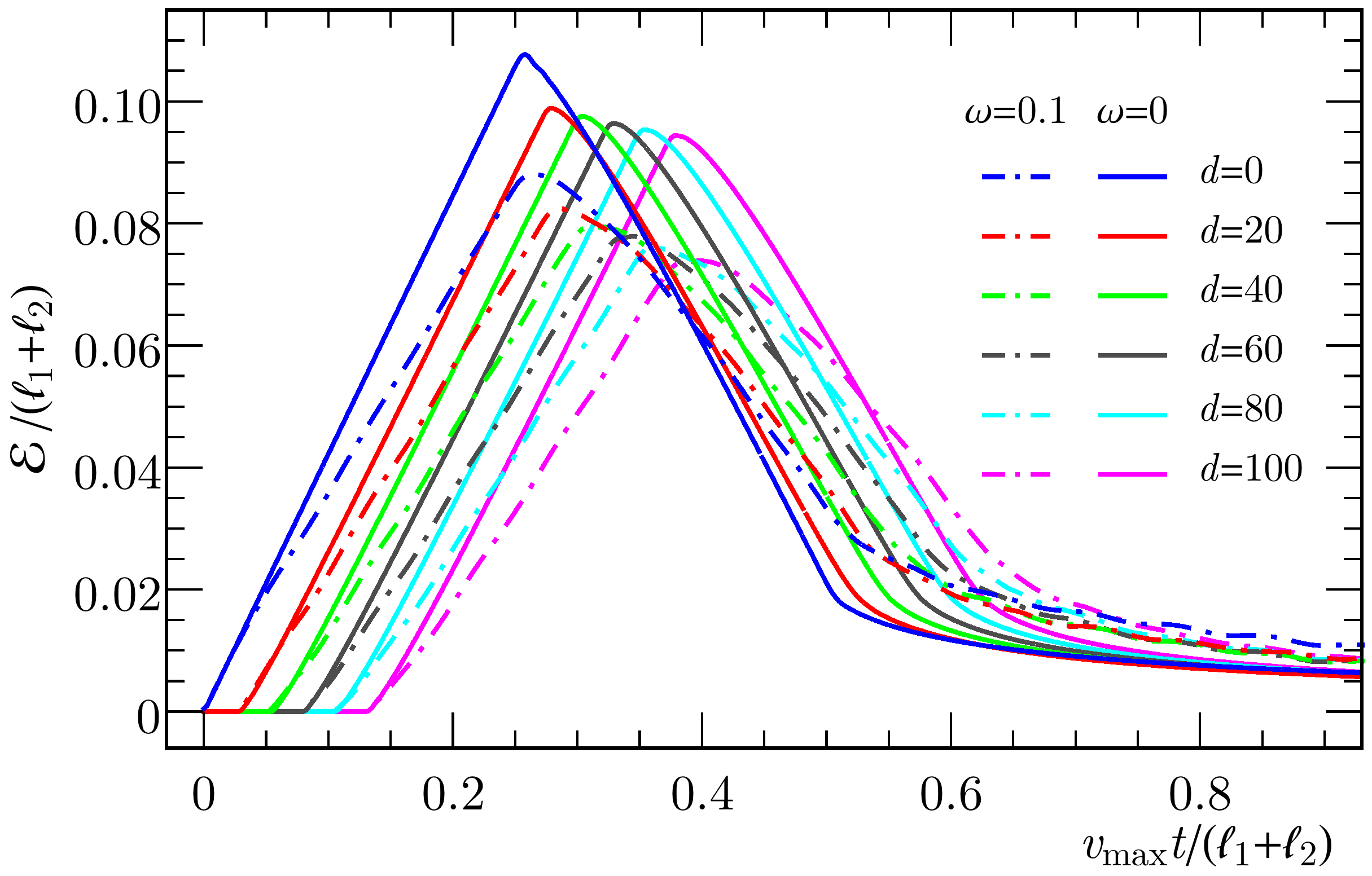}
\caption{
Logarithmic negativity $\mathcal{E}$ of disjoint intervals after global quenches with different evolution Hamiltonians. 
Here $L=5000$, $\ell_1=\ell_2=200$ and $\omega_0=1$, while for $t>0$ we have $\omega=0$ or $\omega=0.1$. 
The velocity $v_{\rm max}$ is given by Eq.  (\ref{vmax def}).
}
\label{fig N2 massive}
\end{figure}

\subsection{Massive evolution}

In this section we briefly discuss what happens when the time evolution is governed by a massive Hamiltonian.
This is elucidated with an example in Fig.~\ref{fig N2 massive} where we report and compare 
critical ($\omega=0$) and noncritical ($\omega=0.1$) evolution of the logarithmic negativity always starting from $\omega_0=1$.
The data are reported against $v_{\rm max}t$ ($v_{\rm max}$ is given by Eq.~(\ref{vmax def})).
Also in this case, the data are perfectly compatible with the quasi-particle picture, but 
we notice an interesting effect. 
The slope of the negativity changes as a function of time as a consequence of 
the entanglement carried by slower quasi-particles which in the case of the non-critical evolution 
have a larger weight because of the non-monotonicity of $v_k$ in Eq.~(\ref{vk}). 
However, also this phenomenon does not come as a surprise and indeed it was already 
observed for the entanglement entropy of a single interval following a quench in the Ising/XY chain \cite{fc-08,ds-11}.

\begin{figure}[t]
\vspace{.3cm}
\hspace{-.4cm}
\includegraphics[width=1.0\textwidth]{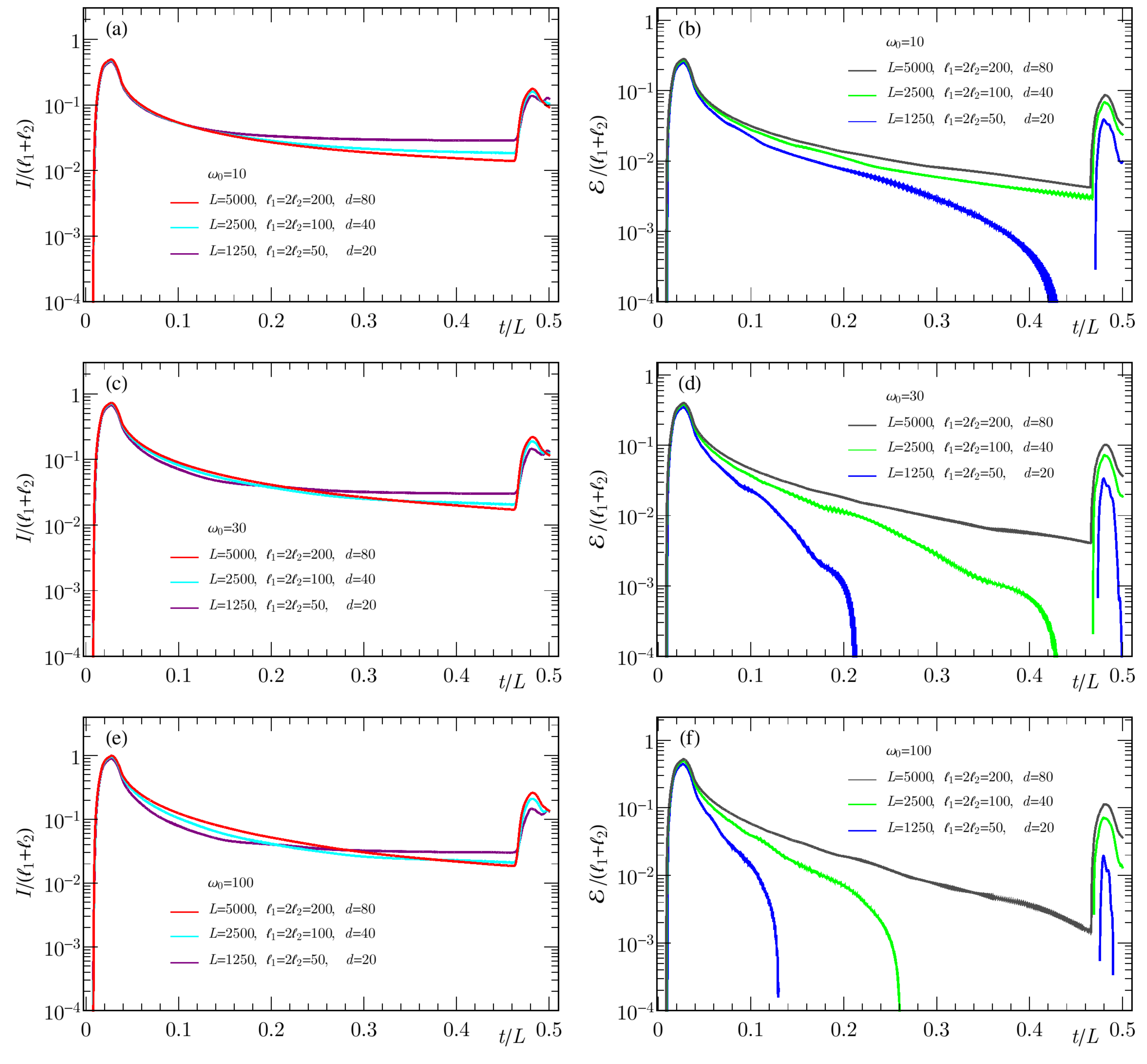}
\caption{
Mutual information (left panels) and logarithmic negativity (right panels) for disjoint intervals with different lengths ($\ell_1=2\ell_2$). 
Here $\omega=0$ and different $\omega_0$ are chosen: $\omega_0=10$,  $\omega_0=30$  and $\omega_0=100$ 
(top, middle and bottom panels respectively). 
The sudden death of the logarithmic negativity happens for later time as $L$ increases and as $\omega_0$ decreases.
}
\label{fig sudden death}
\end{figure}

\subsection{Sudden death of entanglement}

A final important feature of the logarithmic negativity $\mathcal{E}$ is the so called {\it sudden death} of entanglement.
The phenomenon has been first introduced for other entanglement measures \cite{ye-09}, but it has been observed 
also for the negativity. 
For example, in the case of the harmonic chain at thermal equilibrium at some finite $T$, it consists into the exact vanishing 
of $\mathcal{E}$ for temperatures larger than a critical value as discussed also in \cite{us-neg-T,aw-08,fcga-08}, where it was 
emphasised its lattice nature and its absence in the quantum field theory description of the continuum limit. 
In the case of quench, we have already seen that the entanglement and the mutual information of two intervals 
are vanishing for large enough time in the CFT, but this is an independent phenomenon compared to the sudden death.

In order to show the true sudden death of $\mathcal{E}$ after a global quench, we report in Fig.~\ref{fig sudden death} 
the negativity and we compare it to the mutual information for several initial frequencies 
$\omega_0=10$, $\omega_0=30$ and $\omega_0=100$ and for several configurations of the  intervals. 
It is evident that in all cases, while the mutual information stays finite at any time, the negativity $\mathcal{E}$
drops suddenly to zero after some time. 
Thus, the sudden death is a peculiarity of the entanglement and it is not reflected by the correlations (quantified by the 
mutual information) which are always a smooth function of the time. 
Let us now discuss how this phenomenon depends on the various parameters. 
First, we notice that increasing the system size and keeping the ratios between the various lengths fixed, 
the sudden death time increases and, when the system and the subsystems are large enough, 
the sudden death does not occur anymore. 
This agrees with the expectation that the sudden death is a lattice effect. 
Second we observe that the sudden death depends also on the initial frequency $\omega_0$:
increasing $\omega_0$, the sudden  death time decreases.
Finally, it is worth noticing that, when the revival takes place, the entanglement appears again, but this is not at all 
surprising.

\section{Conclusions}
\label{concl}

We studied the evolution of the entanglement negativity following a quantum quench.
We considered the case of a conformal evolution starting from a boundary state. 
First,  for the sake of completeness, we reviewed the results 
for the entanglement entropy and the mutual information of an arbitrary number of (adjacent or disjoint) intervals 
within the path integral approach of Refs.~\cite{cc-05-quench,cc-06,cc-07-quench}.
Then we moved to the calculation of the negativity between two adjacent and disjoint intervals
which are respectively given by Eqs.~(\ref{neg 2adj t-dep cft}) and (\ref{logneg N2disj cft t-dep}).
The applicability and generality of these results have been checked against exact numerical 
calculations for the same quantities in the harmonic chain. 
One of the main results of this study is that the quasi-particle picture \cite{cc-05-quench} 
for the time dependence of the entanglement after a global quantum quench applies also to the 
negativity between two intervals: this is a remarkable and non-trivial property.
We also highlight two peculiar lattice effects: the late birth of entanglement and its sudden death.
The former one consists in the fact that the negativity starts growing slightly after the time predicted by 
the quasi-particle picture, a delay which vanishes in the continuum limit. 
The latter one, instead, is well known and concerns  the exact vanishing of the negativity after some 
given large time (but before revivals take place). 
We investigated how these effects depend on the quench parameters. 

While we draw a quite complete picture for the evolution of the negativity after a quantum quench,
there are still some questions deserving further investigations. 
First, it would be very interesting to obtain analytical forms in the harmonic chain for the evolution of the various entanglement measures 
(entropy, mutual information, and negativity) in analogy to what done in Refs.~\cite{fc-08,fc-10,bkc-14} for the Ising model 
(but only for entropy and mutual information). 
However, we should stress that there are still no analytic results for the entanglement entropy of bosonic  models 
even in the ground state, in contrast with the many results available for free fermion thanks to 
Toeplitz matrix techniques \cite{jk-04,ce-10,fc-11,cmv-12}.
Obtaining such analytic predictions is a prerequisite in order to tackle the quench problem.


\section*{Acknowledgments}

ET thanks  Hong Liu  and Luca Tagliacozzo for useful discussions and
is grateful to the Center for Theoretical Physics at MIT for warm hospitality during part of this work. 
PC and ET acknowledge the financial  support by the ERC under  Starting Grant  279391 EDEQS.


\end{document}